\documentclass[12pt]{article}
\usepackage{amssymb}

\def\hybrid{\topmargin 0pt      \oddsidemargin 0pt
        \headheight 0pt \headsep 0pt
        \voffset=-0.5cm
        \hoffset=-0.25in
        \textwidth 6.75in
        \textheight 9.5in       
        \marginparwidth 0.0in
        \parskip 5pt plus 1pt   \jot = 1.5ex}
\catcode`\@=11
\def\marginnote#1{}

\newcount\hour
\newcount\minute
\newtoks\amorpm
\hour=\time\divide\hour by60 \minute=\time{\multiply\hour by60
\global\advance\minute by-\hour}
\edef\standardtime{{\ifnum\hour<12 \global\amorpm={am}%
        \else\global\amorpm={pm}\advance\hour by-12 \fi
        \ifnum\hour=0 \hour=12 \fi
        \number\hour:\ifnum\minute<10 0\fi\number\minute\the\amorpm}}
\edef\militarytime{\number\hour:\ifnum\minute<10 0\fi\number\minute}

\def\draftlabel#1{{\@bsphack\if@filesw {\let\thepage\relax
   \xdef\@gtempa{\write\@auxout{\string
      \newlabel{#1}{{\@currentlabel}{\thepage}}}}}\@gtempa
   \if@nobreak \ifvmode\nobreak\fi\fi\fi\@esphack}
        \gdef\@eqnlabel{#1}}
\def\@eqnlabel{}
\def\@vacuum{}
\def\draftmarginnote#1{\marginpar{\raggedright\scriptsize\tt#1}}
\def\draftlabel#1{{\@bsphack\if@filesw {\let\thepage\relax
   \xdef\@gtempa{\write\@auxout{\string
      \newlabel{#1}{{\@currentlabel}{\thepage}}}}}\@gtempa
   \if@nobreak \ifvmode\nobreak\fi\fi\fi\@esphack}
        \gdef\@eqnlabel{#1}}
\def\@eqnlabel{}
\def\@vacuum{}
\def\draftmarginnote#1{\marginpar{\raggedright\scriptsize\tt#1}}

\def\draft{\oddsidemargin -.5truein
        \def\@oddfoot{\sl preliminary draft \hfil
        \rm\thepage\hfil\sl\today\quad\militarytime}
        \let\@evenfoot\@oddfoot \overfullrule 3pt
        \let\label=\draftlabel
        \let\marginnote=\draftmarginnote
   \def\@eqnnum{(\theequation)\rlap{\kern\marginparsep\tt\@eqnlabel}%
\global\let\@eqnlabel\@vacuum}  }

\def\numberbysection{\@addtoreset{equation}{section}
        \def\theequation{\thesection.\arabic{equation}}}

\def\underline#1{\relax\ifmmode\@@underline#1\else
        $\@@underline{\hbox{#1}}$\relax\fi}

\def\titlepage{\@restonecolfalse\if@twocolumn\@restonecoltrue\onecolumn
     \else \newpage \fi \thispagestyle{empty}\c@page\z@
        \def\thefootnote{\fnsymbol{footnote}} }

\def\endtitlepage{\if@restonecol\twocolumn \else  \fi
        \def\thefootnote{\arabic{footnote}}
        \setcounter{footnote}{0}}  

\numberbysection \hybrid

\newcounter{mo}

\newcommand{\tr}{{\rm tr}}
\newcommand{\ti}[1]{\tilde{#1}}

\newcommand{\si}{\sigma}

\newcommand{\vf}{\varphi}
\newcommand{\al}{\alpha}
\newcommand{\ka}{\kappa}
\newcommand{\be}{\beta}
\newcommand{\ga}{\gamma}
\newcommand{\om}{\omega}
\newcommand{\vth}{\vartheta}

\newcommand{\bS}{{\mathbb{S}}}

\newcommand{\Mat}{ {\rm Mat}(N,\mathbb C) }
\newcommand{\Matm}{ {\rm Mat}(M,\mathbb C) }
\newcommand{\Matnm}{ {\rm Mat}(NM,\mathbb C) }

\newcommand{\mC}{\mathbb C}
\newcommand{\mZ}{\mathbb Z}

\def\SLN{{\rm SL}(N, {\mathbb C})}

\newtheorem{predl}{Proposition}[section]

\newtheorem{rem}{Remark}

\def\beq{\begin{equation}}
\def\eq{\end{equation}}
\def\p{\partial}

\newcommand{\mats}[4]{\left(\begin{array}{cc}{#1}&{#2}\\ {#3}&{#4}
\end{array}\right)}

\def\res{\mathop{\hbox{Res}}\limits}

\begin{document}

\setcounter{page}{1}

\date{}
\date{}
\vspace{50mm}

\begin{flushright}
 ITEP-TH-14/15\\
\end{flushright}
\vspace{0mm}

\begin{center}
\vspace{0mm}
 {\LARGE{Noncommutative extensions of elliptic integrable}}
 \\ \vspace{4mm}
 {\LARGE{Euler-Arnold tops and Painleve VI equation}}
\\
\vspace{14mm} {\large {A. Levin}\,$^{\flat\,\sharp}$ \ \ \ \ {M.
Olshanetsky}\,$^{\sharp\,\ddagger\,\natural}$
 \ \ \ {A. Zotov}\,$^{\diamondsuit\, \sharp\, \natural}$ }\\
 \vspace{10mm}

 \vspace{2mm} $^\flat$ -- {\small{\sf 
 NRU HSE, Department of Mathematics,
 Myasnitskaya str. 20,  Moscow,  101000,  Russia}}\\
 \vspace{2mm} $^\sharp$ -- {\small{\sf 
 ITEP, B. Cheremushkinskaya str. 25,  Moscow, 117218, Russia}}\\
 \vspace{2mm} $^\natural$ -- {\small{\sf MIPT, Inststitutskii per.  9, Dolgoprudny,
 Moscow region, 141700, Russia}}\\
\vspace{2mm} $^\ddagger$ -- {\small{\sf
IITP (Kharkevich Institute) RAS,
Bolshoy Karetny per. 19, Moscow, 127994,  Russia}}\\
\vspace{2mm} $^\diamondsuit$ -- {\small{\sf Steklov Mathematical
Institute of Russian Academy of Sciences,\\ Gubkina str. 8, Moscow,
119991,  Russia}}
\end{center}

\begin{center}\footnotesize{{\rm E-mails:}{\rm\ \
 alevin@hse.ru,\  olshanet@itep.ru,\  zotov@mi.ras.ru}}\end{center}

 \begin{abstract}
 In this paper we suggest generalizations of elliptic integrable tops
 to matrix-valued  variables. Our consideration is based on
 $R$-matrix description which provides Lax pairs in terms of quantum
 and classical $R$-matrices. First, we prove that for relativistic
 (and non-relativistic) tops such Lax pairs with spectral parameter
 follow from the associative Yang-Baxter equation and its
 degenerations. Then we proceed to matrix extensions of the models and find
 out that some additional constraints are required for their construction.
 We describe a matrix version of ${\mathbb Z}_2$ reduced
 elliptic top and verify that the latter
 constraints are fulfilled in this case. The construction of matrix extensions is naturally generalized to
 the monodromy preserving equation. In this way we get matrix extensions of
 the Painlev\'e VI equation and its multidimensional analogues
 written in the form of non-autonomous elliptic tops.
 Finally, it is mentioned that
 the matrix valued variables can be replaced by elements of
 noncommutative associative algebra.
 In the end of the paper we also describe
 special elliptic Gaudin models which can be considered as matrix
 extensions of the (${\mathbb Z}_2$ reduced) elliptic top.
 \end{abstract}

\newpage

{\small{

\tableofcontents

}}



\section{Introduction and summary}
\setcounter{equation}{0}

Noncommutative generalizations of integrable systems have long
history started from the non-abelian generalization of the Toda
model proposed by A. Polyakov\footnote{See Appendix by I. Krichever
in paper \cite{Dubrovin81}.}. The incomplete list of papers devoted
to this subject is \cite{EGR,MS,OS,Odess,Konts} and references
therein. The generalization means a passage in the equations of
motion to the variables taking values in associative algebras,
possibly with additional structures. This can be treated as
quantization of the original system. On the other hand, in this way
one can pass from the classical finite-dimensional Hamiltonian
systems to corresponding field theories. Our construction of the
noncommutative integrable systems is based on the associative
Yang-Baxter equation for (quantum) $R$-matrices. We will show that
existence of this equation governs integrability of the related
top-like system. Then it is mentioned that any such $R$-matrix can
be simply generalized to the one corresponding to matrix-valued
extension of the initial top. Finally, we prove that this extension
is indeed integrable under addition reduction procedure.

In this paper we describe a noncommutative generalization of
integrable Euler-Arnold tops related to the group SL$(N,\mC)$. The
simplest example of the latter is given by the Euler top:
 \beq\label{e01}
 \begin{array}{c}
  \displaystyle{
\dot S=[S,J(S)]\,,
 }
 \end{array}
 \eq
 \beq\label{e02}
 \begin{array}{c}
  \displaystyle{
S=\sum\limits_{\al=1}^3\frac{1}{2\imath}\,\sigma_\al
S_\al\,,\quad\quad
J(S)=\sum\limits_{\al=1}^3\frac{1}{2\imath}\,\sigma_\al S_\al
J_\al\,,
 }
 \end{array}
 \eq
where $\si_\al$ are the Pauli matrices, $\imath=\sqrt{-1}$,
$J_1,J_2,J_3$ - arbitrary constants (inverse components of inertia
tensor written in principle axes) and $(S_1,S_2,S_3)$ -- the
dynamical variables (components of the angular momentum vector). The
model is Hamiltonian. Its phase space is parameterized by the
$S_\al$ variables treated as coordinates on ${\rm su}^*(2)$ Lie
coalgebra, where the Poisson-Lie structure is defined:
 \beq\label{e03}
 \begin{array}{c}
  \displaystyle{
\{S_\al,S_\be\}=\varepsilon_{\al\be\ga}S_\ga\,,\quad\quad
H=\frac12\,\sum\limits_{\al=1}^3 J_\al S_\al^2\,.
 }
 \end{array}
 \eq
 The Hamiltonian equations ${\dot S}_\al=\{H,S_\al\}$ are equivalent
 to (\ref{e01}).
 In what follows
we deal with the complexified version of the Euler equation and its
generalizations, i.e. $S_\al\in\mC$, $J_\al\in\mC$ and ${\rm
su}^*(2)$ is replaced by ${\rm sl}^*(2,\mC)$.

The Euler-Arnold generalizations of (\ref{e01}) correspond to higher
rank Lie algebras (or groups). It means that $S=\sum_\al S_\al
T_\al$, where $\{T_\al\}$ -- some basis in the Lie algebra
${\mathfrak g}$. Such type dynamical systems were introduced by
Arnold \cite{Arnold}, and were shown to be Liouville integrable in
some particular cases \cite{Di,MF,Ma}.
 We focus on elliptic integrable systems which appeared originally
 for many-body systems of Calogero-Moser type \cite{OP}. The
 construction of its solutions \cite{Krich1} requires the Lax pair with spectral
 parameter $z$ living  on an elliptic curve $\Sigma_\tau=\mC/\mZ\oplus\tau\mZ$ with moduli
 $\tau$,
 Im$\tau>0$. For the top like systems such Lax pairs were constructed
 for continuous and discrete XYZ models by E. Sklyanin \cite{SklyaninLL,Skl1} and
  then were generalized to the Gaudin type
 models and to higher rank cases \cite{Cherednik2,STSR,Nekrasov}
 using the Belavin-Drinfeld elliptic $r$-matrix \cite{BD}. Later
 both types of elliptic models (the many-body systems and the elliptic tops) were
 unified \cite{LOZ} by the Symplectic Hecke correspondence
 (the classical analogue of the IRF-Vertex
 correspondence \cite{Baxter_IRF}).
 Classification of general elliptic models  including
 those of mixed types for simple Lie groups can be found in \cite{LOSZ}.

\noindent \underline{{\bf Elliptic ${\rm sl}_N$ top}} is a
generalization of the Euler one (\ref{e01}) for the case $S\in\Mat$:
 \beq\label{e04}
 \begin{array}{c}
  \displaystyle{
\dot S=[S,J(S)]\,,\quad S=\sum\limits_{i,j=1}^N
E_{ij}S_{ij}=\sum\limits_{\al\in\,\mZ_N\times\mZ_N;\, \al\neq
0}T_\al S_\al\,,
 }
 \end{array}
 \eq
 \beq\label{e0400}
 \begin{array}{c}
  \displaystyle{
  J(S)=\sum\limits_{\al\neq 0}T_\al S_\al J_\al\,,\quad J_\al=-E_2(\om_\al)\,,\quad
 \om_\al=\frac{\al_1+\al_2\tau}{N}\,,
 }
 \end{array}
 \eq
where $\al\neq 0$ is a short notation for $\al=(\al_1,\al_2)\neq
(0,0)$, the set $\{T_a\}$ is a higher rank analogue of the Pauli
matrices basis in $\Mat$ (see (\ref{e904})-(\ref{e9061})), and $E_2$
is the second Eisenstein elliptic function (\ref{e9081}). The
absence of $\al=0$ term in (\ref{e04}) means that $\tr S=0$.

 The (inverse) inertia
tensor $J$ depends on only one complex parameter -- the moduli
$\tau$ of elliptic curve. In fact, one can multiply $J(S)$ by
arbitrary constant and shift all the components $J_\al$ by another
one constant (the latter does not effect equations of motion). Thus
we have three parameters, and in this sense the elliptic ${\rm
sl}(2,\mC)$ top (in this case $\{\om_\al\}$ is the set of
half-periods $\{0,\tau/2,1/2+\tau/2,1/2\}$) coincides with the
complexified Euler top.

The Lax equations
 \beq\label{e040}
 \begin{array}{c}
  \displaystyle{
{\dot L}(z,S)=[L(z,S),M(z,S)]
 }
 \end{array}
 \eq
written for the Lax pair with the spectral parameter $z$
 \beq\label{e041}
 \begin{array}{c}
  \displaystyle{
L(z)=\sum\limits_{\al\neq 0}T_\al S_\al\vf_\al(z,\om_\al)\,,\quad
M(z)=\sum\limits_{\al\neq 0}T_\al S_\al f_\al(z,\om_\al)
 }
 \end{array}
 \eq
are equivalent to (\ref{e04})-(\ref{e0400}) identically in $z$. The
functions entering (\ref{e041}) are given in (\ref{e9102}),
(\ref{e9103}). Let us also write down equations of motion
(\ref{e04}) in components $S_\al$ (i.e. equations as coefficient
behind $T_\al$)\footnote{In $N=2$ case equations (\ref{e0411})
coincide with (\ref{e01}) up to redefinition (\ref{e9061}) an the
factor $1/2\imath$ as in (\ref{e02}).}:
 \beq\label{e0411}
 \begin{array}{c}
  \displaystyle{
{\dot S}_\al=\sum\limits_{\be,\ga:\,\be+\ga=\al}
(\ka_{\be,\ga}-\ka_{\ga,\be})S_\be S_\ga J_\ga\,,\ \al\neq 0\,,
 }
 \end{array}
 \eq
where $\ka_{\be,\ga}$ are structure constants defined by relations
$T_\be T_\ga=\ka_{\be,\ga}T_{\be+\ga}$ (\ref{e905}).

\noindent \underline{{\bf Relativistic elliptic ${\rm gl}_N$ top}}
is a deformation of (\ref{e04})-(\ref{e0400}). It generalizes the
non-relativistic top in the same way as the (elliptic)
Ruijsenaars-Schneider model \cite{Ruijs1} generalizes the
Calogero-Moser model. The Lax equation (\ref{e040}) is written for
the Lax pair
 \beq\label{e042}
 \begin{array}{c}
  \displaystyle{
{L^\eta}(z)=\sum\limits_{\al} T_\al S_\al \vf_\al^\eta(z)\,,\quad
M^\eta(z)=-\sum\limits_{\al\neq 0}T_\al S_\al\vf_\al(z,\om_\al)\,,
\quad S\in\Mat\,,
 }
 \end{array}
 \eq
where $\eta$ is the deformation parameter and $\{\vf_a^\eta(z)\}$ is
the set of functions (\ref{e910}). It provides the equations of
motion
 \beq\label{e043}
 \begin{array}{c}
  \displaystyle{
\dot S=[S,J^\eta(S)]\,,
 }
 \end{array}
 \eq
 \beq\label{e044}
 \begin{array}{c}
  \displaystyle{
 J^\eta(S)=\sum\limits_{\al\neq 0}T_\al S_\al J_\al^\eta\,,\quad
 J_\al^\eta=E_1(\eta+\om_\al)-E_1(\om_\al)\,,
 }
 \end{array}
 \eq
where $E_1$ -- is the first Eisenstein function (\ref{e908}). For
the rank 1 matrix $S$ this model is gauge equivalent to the elliptic
Ruijsenaars-Schneider model. In the limit $\eta\rightarrow 0$
(\ref{e043})-(\ref{e044}) turns into (\ref{e04})-(\ref{e0400}).
Similarly to (\ref{e0411}) we have the following equations of motion
written in components $S_\al$:
 \beq\label{e0412}
 \begin{array}{c}
  \displaystyle{
{\dot S}_\al=\sum\limits_{\be,\ga:\,\be+\ga=\al}
(\ka_{\be,\ga}-\ka_{\ga,\be})S_\be S_\ga J_\ga^\eta\,,\ \al\neq
0;\quad {\dot S}_0=0\,.
 }
 \end{array}
 \eq
The relativistic top has also $\eta$-independent description, which
at the level of equations of motion coincides with the
non-relativistic one\footnote{This is because the relativistic top
is a quasi-classical version of one site spin chain, and due to the
fact that the elliptic top admits bihamiltonian structure consisting
of linear and quadratic Poisson $r$-matrix structures. See details
in \cite{LOZ8,KLO}.}. Substitution
 \beq\label{e045}
 \begin{array}{c}
  \displaystyle{
 S_\al\rightarrow S_\al\vf_\al(\eta,\om_\al)\ \ \hbox{for}\ \al\neq
 0\
 \hbox{and}\
 S_0\rightarrow S_0
 }
 \end{array}
 \eq
transforms (\ref{e0412}) into (\ref{e0411}). It can be easily
verified if one represents $J^\eta_\ga$ using (\ref{e9071}) as
$J_\ga^\eta=f_\ga(\eta,\om_\ga)/\vf_\ga(\eta,\om_\ga)$. Then $\eta$
is cancelled out from equations of motion in the same way as
spectral parameter $z$ is cancelled out from the Lax equations
$(\ref{e040})$ providing (\ref{e0411}).


 \noindent \underline{{\bf R-matrix formulation.}}
{The (non)relativistic classical tops can be described in terms of
quantum $R$-matrices} \cite{LOZ8}. In the elliptic case\footnote{See
\cite{LOZ81} and references therein for the rational and
trigonometric cases.} we deal with the Baxter-Belavin ${\rm GL}_N$
$R$-matrix \cite{Belavin} written in the form:
 \beq\label{e050}
 \begin{array}{c}
  \displaystyle{
 R^\hbar_{12}(z_1,z_2)= R^\hbar_{12}(z_1-z_2)=
 \sum\limits_{a\in\,{\mathbb Z}_N\times{\mathbb Z}_N} \vf_a^\hbar(z_1-z_2)\, T_a\otimes
 T_{-a}\in \Mat^{\otimes 2}\,,
 }
 \end{array}
 \eq
It satisfies the quantum Yang-Baxter equation
 \beq\label{e051}
 \begin{array}{c}
  \displaystyle{
R_{12}^\hbar(z_1,z_2)R_{13}^\hbar(z_1,z_3)R_{23}^\hbar(z_2,z_3)=
R_{23}^\hbar(z_2,z_3)R_{13}^\hbar(z_1,z_3)R_{12}^\hbar(z_1,z_2)
 }
 \end{array}
 \eq
and the unitarity condition which for (\ref{e050}) is as follows:
  \beq\label{e052}
  \begin{array}{c}
  \displaystyle{
R^\hbar_{12}(z_1,z_2) R^\hbar_{21}(z_2,z_1)=
N^2(\wp(N\hbar)-\wp(z_1-z_2))\,\,1\otimes 1\,.
 }
 \end{array}
 \eq
The construction of the (non)relativistic tops uses coefficients of
local expansions near $\hbar=0$ (the classical limit)
  \beq\label{e053}
  \begin{array}{c}
  \displaystyle{
R^\hbar_{12}(z)=\frac{1}{\hbar}\,1\otimes 1+r_{12}(z)+\hbar\,
m_{12}(z)+O(\hbar^2)
 }
 \end{array}
 \eq
and near $z=0$:
  \beq\label{e054}
  \begin{array}{c}
  \displaystyle{
R^\hbar_{12}(z)=\frac{N}{z}\, P_{12}+R_{12}^{\hbar,(0)}+z\,
R_{12}^{\hbar,(1)}+O(z^2)\,,
 }
 \end{array}
 \eq
  \beq\label{e055}
  \begin{array}{c}
  \displaystyle{
r_{12}(z)=\frac{N}{z}\, P_{12}+r_{12}^{(0)}+O(z)\,,
 }
 \end{array}
 \eq
 where $P_{12}$ is the permutation operator (\ref{e9052}).
The coefficient $r_{12}(z)$  from expansion (\ref{e053}) is the
classical Belavin-Drinfeld $r$-matrix \cite{BD} (\ref{e922}).
Explicit
 expressions for the coefficients are given in
 (\ref{e922})-(\ref{e925}).

The elliptic top (\ref{e04})-(\ref{e041}) is formulated in terms of
$R$-matrix data as follows:
  \beq\label{e056}
  \begin{array}{c}
    \displaystyle{
J(S)=\tr_2(m_{12}(0)S_2)\,, \quad S_2=1\otimes S\,,\quad \tr
(S)=0\,,
 }
 \end{array}
 \eq
  \beq\label{e0561}
  \begin{array}{c}
    \displaystyle{
L(z,S)=\tr_2(r_{12}(z)S_2)\,, \quad M(z,S)=\tr_2(m_{12}(z)S_2)\,.
 }
 \end{array}
 \eq
 Similarly, for the relativistic elliptic top (\ref{e042})-(\ref{e044}) we
have:
  \beq\label{e081}
  \begin{array}{c}
    \displaystyle{
J^\eta(S)=\tr_2\left(\left(R_{12}^{\eta,(0)}-r_{12}^{(0)}\right)S_2\right)\,,
 }
 \end{array}
 \eq
  \beq\label{e082}
  \begin{array}{c}
    \displaystyle{
L^\eta(z,S)=\tr_2(R^\eta_{12}(z)S_2)\,, \quad
M^\eta(z,S)=-\tr_2(r_{12}(z){\bar S}_2)\,,
 }
 \end{array}
 \eq
where ${\bar S}$ is a traceless part of $S$. Details can be found in
\cite{LOZ8}. $M^\eta$ has no explicit dependence on $\eta$. We keep
this notation to emphasize that it is the $M$-matrix of the
relativistic model.

 \noindent \underline{{\bf $\mZ_2$ reductions in elliptic tops.}}
 To pass to the noncommutative version of the defined above elliptic tops we will need to impose
 some constraints. They can be described for the elliptic tops as the $\mZ_2$ reduction.
 The idea of reduction provided by some finite group in the classical integrable systems was
 proposed by Aleksander Mikhailov \cite{Mi}. It allows one to construct non-trivial integrable systems
 starting from some trivial or known integrable systems.

 The $\mZ_2$ reduction under consideration is simply written in
 terms of coordinates on the phase space $S_\al$. The corresponding
 constraints are
  \beq\label{e0855}
  \begin{array}{c}
    \displaystyle{
S_\al=S_{-\al}\ \ \hbox{for}\ \hbox{all}\ \al
 }
 \end{array}
 \eq
for non-relativistic top and
  \beq\label{e0877}
  \begin{array}{c}
    \displaystyle{
\frac{S_\al}{\vf_\al(\eta,\om_\al)}=\frac{S_{-\al}}{\vf_{-\al}(\eta,-\om_\al)}\,,
\ \ \hbox{for}\ \hbox{all}\ \al\neq 0
 }
 \end{array}
 \eq
in relativistic case. Some details of the reduction are given in the
Appendix. Let us just mention here that in $N=2$  case (which is the
Euler top (\ref{e01})-(\ref{e03})) the reduction is trivial since
the constraints (\ref{e0855}) and (\ref{e0877}) are identities.
Indeed, $T_\al\equiv T_{-\al}=\si_\al$ and $S_\al\equiv S_{-\al}$.
The arguments $\om_\al$ are half-periods $\tau/2,(\tau+1)/2,1/2$,
therefore, using (\ref{e77})-(\ref{e78}) it is easy to show that
$\vf_\al(\eta,\om_\al)=\vf_{-\al}(\eta,-\om_\al)$.
 As we will see below in the reduced case one can replace commuting
 variables by non commuting.

The classical $r$-matrix structure on the reduced phase space turns
into the classical reflection equation \cite{Skl2}.
 Two important examples of such type reduction were described in
 \cite{Z04} and \cite{LOZ5}. The first one is the BC$_1$ Calogero-Inozemtsev
 model \cite{Inoz} described by equation
  \beq\label{e0871}
\frac{d^2u}{d t^2}=\sum\limits_{a=0}^{3}\nu_a^2\wp'(u+\om_a)\,.
  \eq
The second example is the Zhukovsky-Volterra gyrostat \cite{Zhuk}.
It generalizes the Euler top (\ref{e01}) to non-zero external field
  \beq\label{e0872}
 \p_t S=[S,J(S)]+[S,\nu']\,,
  \eq
where $\nu'=\sum\limits_{\al=1}^3\nu'_\al\sigma_\al$, and
$(\nu'_1\,,\nu'_2\,,\nu'_3)$ plays the role of constant external
field
 (gyrostatic momentum in classical case and magnetic field in
 quantum case).
The Lax pair for (\ref{e0872}) generalizes (\ref{e041}) in the
following way:
 \beq\label{e0873}
 \begin{array}{c}
  \displaystyle{
L^{ZV}(z)=\frac{1}{2\imath}\sum\limits_{\al=1}^3\sigma_\al
\left(S_\al\vf_\al(z,\om_\al)+\frac{\nu'_\al}{\vf_\al(z,\om_\al)}\right)\,,}
\\ \ \\
  \displaystyle{
M^{ZV}(z)=-\frac{1}{2\imath}\sum\limits_{\al=1}^3 \sigma_\al S_\al
\frac{\vf_1(z,\om_1)\vf_2(z,\om_2)\vf_3(z,\om_3)}{\vf_\al(z,\om_\al)}\,.
 }
 \end{array}
 \eq
 The models (\ref{e0871}) and (\ref{e0872}) are gauge equivalent at
 the level of Lax pairs.
 Explicit change of variables $S_a=S_a({\dot u}, u,
 \nu_0,...,\nu_3)$ was obtained in \cite{LOZ5}. The constants
 $\nu'_\al$ from (\ref{e0872}) are linear combinations of $\nu_\al$ from (\ref{e0871})
with $\tau$-dependent coefficients (see (\ref{e0875})-\ref{e0876})).
The fourth (missing) constant in (\ref{e0872}) appears as the value
of (Casimir function) ${\nu'_0}^2=S_1^2+S_2^2+S_3^2$.

 \noindent \underline{{\bf Painlev\'{e} VI equation as non-autonomous
 top.}} The Painlev\'{e} VI equation is the top equation in the hierarchy of the classification of
non-linear ODE of second order possessing the Painlev\'{e} property.
It depends on four constants and
 can be defined as the monodromy preserving condition
 for a linear differential system
with meromorphic coefficients defined on $\mC P^1$.
 Equivalently, it can be formulated in elliptic form
\cite{Pa,M}. Then it takes the form of a non-autonomous version of
the Calogero-Inozemtsev system BC$_1$ (\ref{e0871}):
  \beq\label{e0874}
\frac{d^2u}{d \tau^2}=\sum\limits_{a=0}^{3}\nu_a^2\wp'(u+\om_a)\,,
  \eq
  while the monodromy preserving condition is of the form:
  \beq\label{e0870}
 \begin{array}{c}
  \displaystyle{
 \p_\tau L(w)-\frac{1}{2\pi\imath}\,\p_w M(w)=[L(w),M(w)]\,.
 }
 \end{array}
 \eq
Equation (\ref{e0874}) is non-autonomous since $\wp'(u+\om_a)$
depends on moduli $\tau$ in both - explicit (through its dependence
on $\om_a$) and implicit (through definition (\ref{e9081}) of
$\wp$-function) ways.
 Similarly, one can define the non-autonomous version of the Zhukovsky-Volterra
  gyrostat (\ref{e0872})
\beq\label{e0111}
 \begin{array}{c}
  \displaystyle{
 \p_\tau S=[S,J(S)]+[S,\nu']\,.
 }
 \end{array}
 \eq
The latter model is non-autonomous due to $\tau$-dependence of the
components of (inverse) inertia tensor $J_\al$ (\ref{e0400}) and the
$\tau$-dependence entering $\nu'_\al$:
 \beq\label{e0875}
 \begin{array}{c}
  \displaystyle{
 \nu'_\al=c_\al(\tau)\tilde{\nu}_\al\,,\quad c_\al(\tau)=\vf_\al(z,\om_\al)\vf_\al(z-\om_\al,\om_\al)=-\exp(-2\pi
i\,\om_a\p_\tau\om_a)\left(\frac{\vth'(0)}{\vth(\om_a)}\right)^2\,,
 }
 \end{array}
 \eq
 \beq\label{e0876}
 \begin{array}{c}
{\tilde\nu}_0=\frac{1}{2}\left( \nu_0+\nu_1+\nu_2+\nu_3 \right)\,,
\quad {\tilde\nu}_1=\frac{1}{2}\left( \nu_0+\nu_1-\nu_2-\nu_3
\right)\,,
\\
{\tilde\nu}_2=\frac{1}{2}\left( \nu_0-\nu_1+\nu_2-\nu_3 \right)\,,
\quad {\tilde\nu}_3=\frac{1}{2}\left( \nu_0-\nu_1-\nu_2+\nu_3
\right)\,,
 \end{array}
 \eq
where the set of $\nu_a$ consists of $\tau$-independent constants
from (\ref{e0874}). Equations (\ref{e0874}) and (\ref{e0111}) are
again (as in autonomous case) gauge equivalent. The corresponding
change of variables $S_a=S_a({\dot u}, u, \tau,
 \nu_0,...,\nu_3)$ is given in \cite{LOZ5}. In this sense equation (\ref{e0111}) is also
 a form of the Painlev\'e VI equation\footnote{See also \cite{Zot} for interrelations
 between elliptic forms of the Painlev\'{e} VI.}.  The Lax pair generating (\ref{e0111})
through (\ref{e0870}) is (almost) the same as in the autonomous
case:
 \beq\label{e0878}
 \begin{array}{c}
  \displaystyle{
L^{PVI}(w)=L^{ZV}(w)\stackrel{(\ref{e0875})}{=}\frac{1}{2\imath}\sum\limits_{\al=1}^3\sigma_\al
\left(S_\al\vf_\al(w,\om_\al)+{{\tilde\nu}_\al}{\vf_\al(w-\om_\al,\om_\al)}\right)\,,}
\\ \ \\
  \displaystyle{
M^{PVI}(w)=M^{ZV}(w)+E_1(w)L^{ZV}(w)\,.
 }
 \end{array}
 \eq
It is an example of the so-called classical Painlev\'e-Calogero
correspondence \cite{LO} claiming that properly defined Lax pairs
for elliptic non-relativistic models describe both - integrable
mechanics through the Lax equation (\ref{e040}) and the monodromy
preserving equation through (\ref{e0870}). The proof of this fact is
based on the heat equation (\ref{A.4b}) for the Kronecker function.
In a general (Euler-Arnold) case the heat equation holds for
$R$-matrices (\ref{e053}):
 \beq\label{e0879}
 \begin{array}{c}
  \displaystyle{
 2\pi\imath\,\p_\tau R_{12}^\hbar(z)=\p_z\p_\hbar R_{12}^\hbar(z)\,,
 \quad 2\pi\imath\,\p_\tau r_{12}(z)=\p_z m_{12}(z)\,.
 }
 \end{array}
 \eq
In ${\rm sl}_N$ case  substitution of the Lax pair (\ref{e041}) into
(\ref{e0411}) leads to non-autonomous Euler-Arnold top
 \beq\label{e0414}
 \begin{array}{c}
  \displaystyle{
 \p_\tau S=[S,J(S)]\,,
 }
 \end{array}
 \eq
which can be considered as multidimensional analogue of Painlev\'e
equations.

\vskip4mm \noindent \underline{{\bf Purpose of paper:}}

{\bf 1. Lax equations from associative Yang-Baxter equation.} The
quantum Baxter-Belavin $R$-matrix (\ref{e050})-(\ref{e052}) can be
interpreted as matrix generalization of the Kronecker function
(\ref{e907}) \cite{Pol,LOZ9,LOZ10,LOZ11}. Similarly to this scalar
function $R$-matrix satisfies relations which are matrix analogues
of the elliptic function identities and properties. The most
important for our purposes (see also Appendix) are:

 \begin{itemize}
\item associative Yang-Baxter equation \cite{Pol} (analogue of the Fay
identity (\ref{e909})):
 \beq\label{e10}
 \begin{array}{c}
  \displaystyle{
 R^\hbar_{12}
 R^{\eta}_{23}=R^{\eta}_{13}R_{12}^{\hbar-\eta}+R^{\eta-\hbar}_{23}R^\hbar_{13}\,,\
 \ \ R^\hbar_{ab}=R^\hbar_{ab}(z_a-z_b)\,,
 }
 \end{array}
 \eq
 \item skew-symmetry (analogue of $\phi(\hbar,z)=-\phi(-\hbar,-z)$ and $E_1(z)=-E_1(-z)$):
 \beq\label{e11}
 \begin{array}{c}
  \displaystyle{
 R^{\hbar}_{12}(z)=-R^{-\hbar}_{21}(-z)\,,\ \ \ r_{12}(z)=-r_{21}(-z)\,,\ \ \
 m_{12}(z)=m_{21}(-z)\,,
  }
 \end{array}
 \eq
 %
 \end{itemize}

In Section \ref{sect2} it is shown that Lax equations with Lax pairs
of the relativistic (\ref{e082}) or non-relativistic (\ref{e0561})
top are equivalent to equations of motion (\ref{e043}) or
(\ref{e04}) with the inverse inertia tensors (\ref{e081}) or
(\ref{e056}) respectively. We do not explicitly use the elliptic
function identities. Our derivation is valid for
any $R$-matrices (\ref{e050})-(\ref{e055}) satisfying also
(\ref{e10}), (\ref{e11}).

{\bf 2. Matrix extensions of tops.} A direct meaning of a matrix
extension is that (the scalar, $\mC$-valued) variables of a model
are replaced by noncommutative $\Matm$ matrices. See examples in
\cite{Odess}. When $M=1$ we come back to initial system. Matrix
extension can be thought of as noncommutative version of a model. It
is then described by noncommutative (double) Poisson brackets
\cite{Konts}. Appearance of matrix variables provides also
additional ${\rm GL}_M$ symmetry: this group acts on all matrix
variables by conjugation. The corresponding Poisson algebra and its
quantization was studied in \cite{Avan}. The double brackets
formalism is not used in our paper. Our aim is to get equations of
motion for matrix extensions by generalizing the Lax pairs.

The set of variables (or the coordinates on the phase space) in the
elliptic ${\rm gl}_N$ top $\{S_\al\in\mC,\ \al\in\mZ_N\times
\mZ_N\}$ should be replaced by set of matrices
$\{\bS_\al\in\Matm\}$:
 \beq\label{e13}
 \begin{array}{c}
  \displaystyle{
 S_\al\ \rightarrow\
 \bS_\al=\sum\limits_{\ti\al\in\,\mZ_M\times\mZ_M} {\ti T}_{\ti
 \al}\,
 \bS_\al^{\ti \al}\in\Matm\,,\quad \al\in\mZ_N\times \mZ_N\,,
 }
 \end{array}
 \eq
where $\{{\ti T}_{\ti \al}\in \Matm,\ \ti\al\in\,\mZ_M\times\mZ_M\}$
is the basis (\ref{e904}) in $\Matm$. We will use tildes for "matrix
" or "noncommutative" space\footnote{The noncommutativity means that
$\bS_\al\bS_\be\neq \bS_\be\bS_\al$, and we do not imply any
constraints for any $\bS_\al$ inside noncommutative space.}. It
becomes "scalar" or "commutative" when $M=1$. The space $\Mat$ is an
auxiliary space. It coincides with the matrix space of Lax equations
(\ref{e040}) or matrix form of equations of motion (\ref{e041}) of
initial (scalar) models.

A natural way to get generalizations of the construction of Lax
pairs (\ref{e082}), (\ref{e0561}) to matrix-valued variables is to
consider the following ${\rm Mat}(NM,\mC)^{\otimes 2}$-valued
$R$-matrix:
 \beq\label{e14}
 \begin{array}{c}
  \displaystyle{
R^\eta_{12,\ti 1\ti 2}(z)=R^\eta_{12}(z)\otimes {\ti P}_{\ti 1 \ti
2}\,,
 }
 \end{array}
 \eq
where $R^\eta_{12}(z)$ is the same ${\rm Mat}(N,\mC)^{\otimes
2}$-valued $R$-matrix in  auxiliary space as in (\ref{e082}), while
${\ti P}_{\ti 1 \ti 2}$ is the permutation operator in
noncommutative space. It is easy to see that $R^\eta_{12,\ti 1\ti
2}(z)$ is indeed $R$-matrix in the sense of quantum Yang-Baxter
equation (\ref{e051}) and unitarity condition (\ref{e052}).
Moreover, it satisfies the associative Yang-Baxter equation
(\ref{e10}) as well. However it has a different to (\ref{e053})
classical limit (it starts not from $\hbar^{-1}\,1_{NM}\times
1_{NM}$). For this reason the general construction of the Lax pairs
does not work for matrix extensions in the same way as in scalar
case. To overcome this problem additional constraints are required.
The first one is that
 \beq\label{e15}
 \begin{array}{c}
  \displaystyle{
\bS_0=1_M\, S_0\,,
 }
 \end{array}
 \eq
i.e. matrix extension of $S_0$ variable should be also scalar. This
condition obviously needs to be preserved by dynamics (equations of
motion). The latter provides another constraint. With these
constraints the generalization of construction of Lax pairs works
for $R$-matrix (\ref{e14}) and provides equations of motion
 \beq\label{e16}
  \begin{array}{c}
  \displaystyle{
 {\dot \bS_{1\ti 1}}=[\bS_{1\ti 1},J_1(\bS_{1\ti 1})]\,,
 }
 \end{array}
 \eq
 where the inertia tensor $J$ acts in auxiliary space only.

We will show that the above mentioned constraints are fulfilled for
matrix extensions of $\mZ_2$ reduced elliptic tops. It means that
similarly to (\ref{e0855}) we set $\bS_\al=\bS_{-\al}$. In this case
one obtains
 \beq\label{e17}
 \begin{array}{c}
  \displaystyle{
{\dot \bS}_\al=\sum\limits_{\be,\ga:\,\be+\ga=\al} \Big(
 \kappa_{\be,\ga}\,\bS_\be\bS_\ga - \kappa_{\ga,\be}\,\bS_\ga\bS_\be \Big)\,
 J_\ga\,,\ \al\neq 0\,.
 }
 \end{array}
 \eq
where $J_\ga=-E_2(\om_\ga)$. In scalar case $M=1$ the latter
equations coincide with (\ref{e0411}). The same holds true for the
relativistic top (\ref{e0412}) and $\mZ_2$-reduction constraints
(\ref{e0877}).
\begin{rem}
In this construction one  can replace the algebra $\Matm$  by an
arbitrary associative noncommutative algebra with a well defined
trace functional and the permutation operator acting on the basis of
the algebra. For example, one can take the infinite group of the
quantum torus, its trigonometric and rational degenerations, or
their quasi-classical limits to the algebras of vector fields.
\end{rem}

\begin{rem}
Our results allows one to define the equations of motion in the
Hamiltonian form using double Poisson brackets
 \cite{Ko,VdB,CB} as it was done in \cite{Odess,Avan,Ar}.
 It is straightforward to construct the "classical" $\tilde{r}$-matrix by means the classical $r$-matrix
 (\ref{e053}) and the permutation operator in $\Matm$ and consider the classical reflection
 equation defined by $\tilde{r}$.  It leads the
 double Poisson brackets for the Lax operators $L^\eta(z,\bS)$
 in terms of the $\tilde{r}$-matrix. Furthermore, it is open a way to quantize the noncommutative
 tops by means of the $R$-matrix (\ref{e14}) and the quantum reflection
 equation.
\end{rem}

In the end of the paper we also describe a special elliptic Gaudin
model with equations of motion
  \beq\label{e18}
  \begin{array}{c}
    \displaystyle{
{\dot A}^\al=\sum\limits_{\be+\ga=\al}[A^\be,A^\ga]\,J_\ga\quad
\al\neq 0\,,
 }
 \end{array}
 \eq
 where $\{A^\al\}$ is a set of $N^2-1$ matrices of size $N\times N$ with constraints $A^{\al}=A^{-\al}$.
 Equations (\ref{e18}) reproduce the elliptic top equations of motion
 (\ref{e0411}) via reduction $A^\al\rightarrow T_\al S_\al$ ($S_\al=S_{-\al}$).

 {\bf 3. Matrix extensions of Painlev\'e equations.}
 Finally, we construct the noncommutative generalization of the Painlev\'e VI equation.
 The non-commutative generalizations of the Painlev\'{e} II-IV equations were considered before
 in \cite{CF,BC,OS,RR}. Here we identify the non-commutative Painlev\'e VI equation with the
 non-commutative non-autonomous Zhukovsky-Volterra gyrostat
 (\ref{e0111}):
 \beq\label{e087868}
 \begin{array}{c}
  \displaystyle{
\frac{d}{d\tau}\,\bS_\al=\frac{1}{2}\,(\bS_\be\bS_\ga+\bS_\ga\bS_\be)(E_2(\om_\be)-E_2(\om_\ga))+\bS_\be\nu'_\ga-\bS_\ga\nu'_\be\,,
 }
 \end{array}
 \eq
 These equations takes the form (\ref{e0111}) for $N=1$.
 Our construction allows one to define the Lax pair for the Painlev\'e VI equation using the same Lax operators as for
 the autonomous case.

\paragraph{Acknowledgments.} The work was supported by RFBR grant
15-31-20484 mol$\_$a$\_$ved and by joint project 15-51-52031
HHC$_a$.
The work of A. Levin was partially supported by Department of
Mathematics NRU HSE, the subsidy granted to the HSE by the
Government of the Russian Federation for the implementation of the
Global Competitiveness Program, and by the Simons Foundation.


 \section{Lax pairs from associative Yang-Baxter equation}\label{sect2}
 \setcounter{equation}{0}
In this Section we do not use explicit forms of Lax pairs but only
the properties of the underlying $R$-matrices. Our current purpose
is to show that the Lax equations with the $R$-matrix forms of the
Lax pairs of integrable tops (\ref{e0561}), (\ref{e082}) are
equivalent to equations of motion (\ref{e04}), (\ref{e043}) with the
corresponding inertia tensors (\ref{e056}), (\ref{e081}) due to
additional properties of $R$-matrix (\ref{e10}), (\ref{e11}). Below
we prove these statements for relativistic and non-relativistic tops
separately.
\begin{predl}
Suppose that quantum $R$-matrix entering the Lax pair of the
relativistic top (\ref{e082}) satisfies not only
(\ref{e051})-(\ref{e055}) but also the associative Yang-Baxter
equation (\ref{e10}) and the skew-symmetry property (\ref{e11}).
Then the Lax equations (\ref{e040}) with the Lax pair (\ref{e082})
are equivalent to equations of motion of the relativistic top
(\ref{e043}) with (inverse) inertia tensor $J^\eta$ (\ref{e081}).
\end{predl}

\vspace{-2mm}  \underline{\em{Proof}}:\vskip1mm

\noindent Let us verify that the Lax equations
 \beq\label{e21}
 \begin{array}{c}
  \displaystyle{
 {\dot L}^\eta(z,S)=[L^\eta(z,S),M^\eta(z,S)]
 }
 \end{array}
 \eq
with $L^\eta$ and $M^\eta$ (\ref{e082}) are fulfilled on equations
of motion
 \beq\label{e22}
 \begin{array}{c}
  \displaystyle{
 {\dot S}=[S,J^\eta(S)]\,,\ \
 J^\eta(S)=\tr_2\left(\left(R_{12}^{\eta,(0)}-r_{12}^{(0)}\right)S_2\right)
 }
 \end{array}
 \eq
identically in spectral parameter $z$, i.e.
 \beq\label{e23}
 \begin{array}{c}
  \displaystyle{
 {L}^\eta(z,[S,J^\eta(S)])=[L^\eta(z,S),M^\eta(z,S)]\,,\ \forall\,
 z\,.
 }
 \end{array}
 \eq
 The l.h.s. of (\ref{e23}) is equal to\footnote{The index 1 in (\ref{e24})
 is the number of tensor component corresponding to the matrix space $\Mat$ of equation
 (\ref{e23}). We sometimes omit this index where it is obvious (for
 example,  in l.h.s. of (\ref{e056})-(\ref{e082})).
 The components 2,3 are under trace.
 }
 \beq\label{e24}
 \begin{array}{c}
  \displaystyle{
 {L}_1^\eta(z,[S,J^\eta(S)])=\tr_{2,3}\Big\{R_{12}^\eta(z)\,[S_2,\left(R_{23}^{\eta,(0)}-r_{23}^{(0)}\right)S_3]\Big\}=
 }
 \\ \ \\
   \displaystyle{
 -\tr_{2,3}\Big\{ [R_{12}^\eta(z),R_{23}^{\eta,(0)}]\, S_2S_3\Big\}
  -\tr_{2,3}\Big\{ [R_{13}^\eta(z),r_{23}^{(0)}]\, S_2S_3\Big\}\,,
 }
 \end{array}
 \eq
 where we have used $r_{23}^{(0)}=-r_{32}^{(0)}$ (\ref{e927}).
  To
 simplify the r.h.s. of (\ref{e23}) notice that $\bar S$ in (\ref{e082}) can be replaced by $S$ since
 the scalar part of $M^\eta$ does not give any input to $[L^\eta,M^\eta]$. In fact $\bar S$ in (\ref{e082}) is used
 in order to match the elliptic definition (\ref{e042}).
 This question will become nontrivial in the case of matrix valued variables.

 Let us write down (\ref{e933}) with $z_3=0$, which is a consequence of the associative Yang-Baxter equation (\ref{e10}):
  \beq\label{e25}
  \begin{array}{c}
  \displaystyle{
[R^{\eta}_{13}(z_1),r_{12}(z_1-z_2)] =[R^\eta_{12}(z_1-z_2),
 R^{\eta}_{23}(z_2)]+[R^{\eta}_{13}(z_1),r_{23}(z_2)]
 }
 \end{array}
 \eq
and consider the limit $z_2\rightarrow 0$ (together with renaming
$z_1:=z$):
  \beq\label{e26}
  \begin{array}{c}
  \displaystyle{
[R^{\eta}_{13}(z),r_{12}(z)] =[R^\eta_{12}(z),
 R_{23}^{\eta,(0)}]+[R^{\eta}_{13}(z),r_{23}^{(0)}]-[\p_z R_{12}^\eta(z),NP_{23}]
 }
 \end{array}
 \eq
The simple pole at $z_2=0$ cancel out due to
$[R_{12}^\eta(z),P_{23}]+[R_{13}^\eta(z),P_{23}]=0$ by definition of
permutation operator. From (\ref{e082}) and (\ref{e26}) we conclude
that
  \beq\label{e27}
  \begin{array}{c}
  \displaystyle{
[L^\eta(z,S),M^\eta(z,S)]_1=-\tr_{2,3}\Big\{[R^{\eta}_{13}(z),r_{12}(z)]\,S_2S_3\Big\}=
}
 \\ \ \\
  \displaystyle{
 =-\tr_{2,3}\Big\{[R^\eta_{12}(z),
 R_{23}^{\eta,(0)}]\,S_2S_3\Big\}-\tr_{2,3}\Big\{[R^{\eta}_{13}(z),r_{23}^{(0)}]\,S_2S_3\Big\}\stackrel{(\ref{e24})}{=}
  {L}_1^\eta(z,[S,J^\eta(S)])\,.
 }
 \end{array}
 \eq
 Here we used that
$\tr_{2,3}\Big\{[\p_z R_{12}^\eta(z),NP_{23}]\,S_2S_3\Big\}=0$ since
$[P_{23},S_2S_3]=0$. In this way we finished the proof of
(\ref{e23}) as identity in $z$ on equations of motion (\ref{e22}).
Conversely, (following \cite{LOZ8}) one can easily obtain the
equations of motion (\ref{e22}) from the Lax equations (\ref{e21})
by taking residue of both parts of (\ref{e21}) at $z=0$.
$\blacksquare$

Let us remark that in \cite{LOZ8} we did not prove (\ref{e23}), i.e.
that the Lax equations are identities in spectral parameter on the
equations of motion. Instead, the following indirect argument was
used: we know that (\ref{e23}) holds true in the elliptic case.
Other cases are degenerations of the elliptic one. A degeneration
procedure can be performed at the level of Lax equation as well as
at the level of equations of motion. That is, we used explicit
elliptic formulae to argue that the Lax equations are identities in
$z$. The above given proof (\ref{e21})-(\ref{e27}) is more general.
It does not use any explicit form. It is direct and based on the
associative Yang-Baxter equation only.

Let us now prove a similar statement for non-relativistic top.
\begin{predl}
Suppose that quantum $R$-matrix entering (through expansion
(\ref{e053})) the Lax pair of the non-relativistic top (\ref{e0561})
satisfies not only (\ref{e051})-(\ref{e055}) but also the
associative Yang-Baxter equation (\ref{e10}) and the skew-symmetry
property (\ref{e11}). Then the Lax equations (\ref{e040}) with the
Lax pair (\ref{e0561}) (with $\tr S=0$) are equivalent to equations
of motion of the non-relativistic top (\ref{e04}) with (inverse)
inertia tensor $J$ (\ref{e056}).
\end{predl}

\vspace{-2mm}  \underline{\em{Proof}}:\vskip1mm

\noindent In this case $\tr S=NS_0=0$. Let us verify that the Lax
equations
 \beq\label{e28}
 \begin{array}{c}
  \displaystyle{
 {\dot L}(z,S)=[L(z,S),M(z,S)]
 }
 \end{array}
 \eq
with $L$ and $M$ (\ref{e0561}) are fulfilled on equations of motion
 \beq\label{e29}
 \begin{array}{c}
  \displaystyle{
 {\dot S}=[S,J(S)]\,,\ \
 J(S)=\tr_2\left(m_{12}(0)S_2\right)
 }
 \end{array}
 \eq
identically in spectral parameter $z$, i.e.
 \beq\label{e30}
 \begin{array}{c}
  \displaystyle{
 {L}(z,[S,J(S)])=[L(z,S),M(z,S)]\,,\ \forall\,
 z\,.
 }
 \end{array}
 \eq
 The l.h.s. of (\ref{e30}) equals
 \beq\label{e31}
 \begin{array}{c}
  \displaystyle{
 {L}(z,[S,J(S)])_1=\tr_{2,3}\Big\{ r_{12}(z)[S_2,m_{23}(0)S_3] \Big\}=
  \tr_{2,3}\Big\{ [m_{23}(0),r_{12}(z)]\, S_2S_3 \Big\}\,.
 }
 \end{array}
 \eq
To simplify the r.h.s. of (\ref{e30}) we use (\ref{e934}). Write it
down for $z_3=0$
 $$
 [m_{13}(z_1),r_{12}(z_1-z_2)]=[r_{12}(z_1-z_2),m_{23}(z_2)]+[m_{12}(z_1-z_2),r_{23}(z_2)]+[m_{13}(z_1),r_{23}(z_2)]
 $$
and consider the limit $z_2\rightarrow 0$ (with renaming $z_1:=z$).
The simple pole at $z_2=0$ cancel out due to
$[m_{12}(z),P_{23}]+[m_{13}(z),P_{23}]=0$ and we have:
 \beq\label{e32}
 \begin{array}{c}
  \displaystyle{
 [m_{13}(z),r_{12}(z)]=[r_{12}(z),m_{23}(0)]-[\p_z m_{12}(z),NP_{23}]
  +[m_{12}(z),r_{23}^{(0)}]+[m_{13}(z),r_{23}^{(0)}]\,.
 }
 \end{array}
 \eq
Now we can compute
 \beq\label{e33}
 \begin{array}{c}
  \displaystyle{
 [L(z,S),M(z,S)]_1=\tr_{2,3}\Big\{ [r_{12}(z),m_{13}(z)]\, S_2S_3
 \Big\}\stackrel{(\ref{e32})}{=}\tr_{2,3}\Big\{ [m_{23}(0),r_{12}(z)]\, S_2S_3
 \Big\}=
 }
 \\ \ \\
  \displaystyle{
 \stackrel{(\ref{e31})}{=}{L}(z,[S,J(S)])_1\,.
 }
 \end{array}
 \eq
In the equality via (\ref{e32}) we used that $\tr_{2,3}\Big\{ [\p_z
m_{12}(z),NP_{23}]\, S_2S_3
 \Big\}=0$ due to $[P_{23},S_2S_3]=0$
 and $$\tr_{2,3}\Big\{ \left( [m_{12}(z),r_{23}^{(0)}]+[m_{13}(z),r_{23}^{(0)}] \right) S_2S_3
 \Big\}=0$$
because the expression
$[m_{12}(z),r_{23}^{(0)}]+[m_{13}(z),r_{23}^{(0)}]$ is skew
symmetric with respect to $2\leftrightarrow 3$ due to the property
$r_{23}^{(0)}=-r_{32}^{(0)}$.

Conversely, we can obtain the equations of motion (\ref{e29}) from
the Lax equations (\ref{e28}) by taking the residue at $z=0$ of its
both sides.  $\blacksquare$


\section{Matrix valued tops}\label{sect4}
\setcounter{equation}{0}

In paragraph below we argue why the construction of Section
\ref{sect2} can not be directly generalized to matrix extensions of
the tops models. It appears that matrix variables are not arbitrary
but satisfy some constraints. Then we mention that these constraints
are fulfilled for $\mZ_2$ reduced models and describe their matrix
extensions.

\paragraph{General construction and constraints.} A general idea of matrix extension is to replace scalar
variables $S_\al\in\mC$ by matrix valued variables $\bS_\al\in\Matm$
(\ref{e13}). The initial scalar variables of a top model $S_\al$
were themselves arranged into the matrix valued variable $S=\sum_\al
T_\al S_\al\in\Mat$ (the residue of the Lax matrix). Therefore, we
deal with the following matrix variable:
 \beq\label{e41}
 \begin{array}{c}
  \displaystyle{
 S\in\Mat\ \rightarrow\
 \bS=\bS_{1\ti 1}=\sum\limits_{\al}^N\sum\limits_{\ti\al}^M T_{\al}\otimes{\ti T}_{\ti
 \al}\,
 \bS_\al^{\ti \al}=\sum\limits_\al T_\al\otimes\bS_\al \in\Matnm\,,
 }
 \end{array}
 \eq
where indices $1$, $\ti 1$ stand for $\Mat$ and $\Matm$ tensor
components respectively likewise it is used in $R$-matrix notations.

Recall that the Lax matrix of integrable top was defined as
(\ref{e0561}) $L^\eta(z,S)=\tr_2(R^\eta_{12}(z)S_2)$. The latter
means that for a given $R$-matrix written in standard basis of
$\Mat$ as
 \beq\label{e411}
 \begin{array}{c}
  \displaystyle{
R^\eta_{12}(z)=\sum\limits_{i,j,k,l=1}^N E_{ij}\otimes E_{kl}\,
R_{ijkl}(z,\eta)
 }
 \end{array}
 \eq
the corresponding Lax matrix (\ref{e082}) is of the form:
 \beq\label{e42}
 \begin{array}{c}
  \displaystyle{
L^\eta(z,S)=\sum\limits_{i,j,k,l=1}^N E_{ij} S_{kl}\,
R_{ijlk}(z,\eta)\,.
 }
 \end{array}
 \eq
A natural way to get a matrix generalization is to consider the
following expression:
 \beq\label{e43}
 \begin{array}{c}
  \displaystyle{
R^\eta_{12,\ti 1\ti 2}(z)=R^\eta_{12}(z)\otimes {\ti P}_{\ti 1 \ti
2}=\sum\limits_{i,j,k,l=1}^N\sum\limits_{m,n=1}^M E_{ij}\otimes
E_{kl}\otimes {\ti E}_{mn}\otimes {\ti E}_{nm}\, R_{ijkl}(z,\eta)\,,
 }
 \end{array}
 \eq
where ${\ti E}_{mn}$ is standard basis in $\Matm$ and ${\ti P}_{\ti
1 \ti 2}$ is the permutation operator in $\Matm^{\otimes 2}$.

First, notice that this expression is again a quantum $R$-matrix. It
satisfies the quantum Yang-Baxter equation
 \beq\label{e44}
 \begin{array}{c}
  \displaystyle{
R_{12,\ti 1\ti 2}^\hbar(z_1,z_2)R_{13,\ti 1\ti
3}^\hbar(z_1,z_3)R_{23,,\ti 2\ti 3}^\hbar(z_2,z_3)= R_{23,\ti 2\ti
3}^\hbar(z_2,z_3)R_{13,\ti 1\ti 3}^\hbar(z_1,z_3)R_{12,\ti 1\ti
2}^\hbar(z_1,z_2)
 }
 \end{array}
 \eq
 due to the Yang-Baxter equation for $R_{12}^\eta(z)$ (\ref{e051})
 and  ${\ti P}_{\ti 1 \ti 2}{\ti P}_{\ti 1 \ti 3}{\ti P}_{\ti 2 \ti 3}=
 {\ti P}_{\ti 2 \ti 3} {\ti P}_{\ti 1 \ti 3} {\ti P}_{\ti 1 \ti 2}$.
 The unitarity condition (\ref{e052}) is fulfilled as well:
  \beq\label{e45}
  \begin{array}{c}
  \displaystyle{
R^\hbar_{12,\ti 1\ti 2}(z_1,z_2) R^\hbar_{21,\ti 2\ti 1}(z_2,z_1)=
N^2(\wp(N\hbar)-\wp(z_1-z_2))\,\,1_N\otimes 1_N\otimes 1_M\otimes
1_M\,.
 }
 \end{array}
 \eq
Moreover, $R$-matrix (\ref{e43}) satisfies the associative
Yang-Baxter equation (\ref{e10}):
 \beq\label{e46}
 \begin{array}{c}
  \displaystyle{
 R^\hbar_{12,\ti 1\ti 2}
 R^{\eta}_{23,\ti 2\ti 3}=R^{\eta}_{13,\ti 1\ti 3}R_{12,\ti 1\ti 2}^{\hbar-\eta}
  +R^{\eta-\hbar}_{23,\ti 2\ti 3}R^\hbar_{13,\ti 1\ti 3}\,,\
 \ \ R^\hbar_{ab,\ti a\ti b}=R^\hbar_{ab,\ti a\ti b}(z_a-z_b)
 }
 \end{array}
 \eq
because of (\ref{e10}) and ${\ti P}_{\ti 1 \ti 2}{\ti P}_{\ti 2 \ti
3}={\ti P}_{\ti 1 \ti 3}{\ti P}_{\ti 1 \ti 2}={\ti P}_{\ti 2 \ti
3}{\ti P}_{\ti 1 \ti 3}$. Such type quantum and classical $R$-matrix
structures were considered in \cite{FM} and \cite{Avan}.

Second, similarly to (\ref{e42}) the Lax matrix corresponding to
$R$-matrix (\ref{e43})
  \beq\label{e47}
  \begin{array}{c}
  \displaystyle{
L^\eta(z,\bS)=\tr_{2,\ti 2}( R^{\,\eta}_{12,\ti 1\ti 2}(z)\,
\bS_{2\ti 2} )=\sum\limits_{i,j,k,l=1}^N E_{ij} \otimes \bS_{kl}\,
R_{ijlk}(z,\eta)\,,\ \ \ \bS_{kl}=\sum\limits_{m,n=1}^M
\bS_{kl}^{mn} {\ti E}_{mn}
 }
 \end{array}
 \eq
is exactly the matrix generalization of (\ref{e42}).

Therefore, we could expect to have a direct generalization (to the
matrix case) of the Lax pairs construction via associative
Yang-Baxter equation described in Section \ref{sect2}. However, we
will see that it does not work in the same way. The reason is that
the $R$-matrix (\ref{e43}) do not satisfy the local expansion of the
classical limit (\ref{e053}). Indeed, near $\hbar=0$
  \beq\label{e48}
  \begin{array}{c}
  \displaystyle{
R^\hbar_{12,\ti 1\ti 2}(z)=\frac{1}{\hbar}\,1_N\otimes
1_N\otimes{\ti P}_{\ti 1 \ti 2}+r_{12,\ti 1\ti 2}(z)+O(\hbar)\,,
 }
 \end{array}
 \eq
 i.e. in contrast to (\ref{e053}) the first coefficient of the expansion (\ref{e48}) is not $1_{NM}\otimes
 1_{NM}$. It makes problem in the following way.
The proofs of equivalence of Lax equations and equations of motion
given in Section \ref{sect2} used not the associative Yang-Baxter
equation itself but its degenerations (\ref{e26}) or (\ref{e32})
which appeared from (\ref{e933}). Equation (\ref{e933}) in its turn
was obtained by subtracting (\ref{e932}) from (\ref{e931}). For
$R$-matrix (\ref{e43}) instead (\ref{e931}), (\ref{e932}) we have
 \beq\label{e49}
  \begin{array}{c}
  \displaystyle{
 R^\hbar_{12,\ti 1\ti 2}\,
 R^{\hbar}_{23,\ti 2\ti 3}=R^{\hbar}_{13,\ti 1\ti 3}\,r_{12,\ti 1\ti 2}+r_{23,\ti 2\ti 3}\,R^\hbar_{13,\ti 1\ti 3}-\p_\hbar
 R_{13,\ti 1\ti 3}^\hbar\, {\ti P}_{\ti 1 \ti 2}\,,
 }
 \end{array}
 \eq
 \beq\label{e50}
  \begin{array}{c}
  \displaystyle{
 R^\hbar_{23,\ti 2\ti 3}\,
 R^{\hbar}_{12,\ti 1\ti 2}=R^{\hbar}_{13,\ti 1\ti 3}\,r_{23,\ti 2\ti 3}+r_{12,\ti 1\ti 2}\,R^\hbar_{13,\ti 1\ti 3}-\p_\hbar
 R_{13,\ti 1\ti 3}^\hbar\, {\ti P}_{\ti 2 \ti 3}\,.
 }
 \end{array}
 \eq
Difference between (\ref{e49}) and (\ref{e50}) contains unwanted
term $\p_\hbar
 R_{13,\ti 1\ti 3}^\hbar\, ({\ti P}_{\ti 2 \ti 3}-{\ti P}_{\ti 1 \ti
 2})$ which equals zero in scalar case $M=1$. Then we need to
 require that
 \beq\label{e51}
  \begin{array}{c}
  \displaystyle{
 \tr_{2,3,\ti2,\ti 3} \Big\{ \p_\hbar R_{13,\ti 1\ti 3}^\hbar\, ({\ti
 P}_{\ti 2 \ti 3}-{\ti P}_{\ti 1 \ti 2})\, \bS_{2\ti 2}\, \bS_{3\ti 3}
 \Big\}= \tr_{2,3,\ti2,\ti 3} \Big\{ \p_\hbar R_{13}^\hbar\, ({\ti P}_{\ti 1 \ti 3}{\ti
 P}_{\ti 2 \ti 3}-{\ti P}_{\ti 1 \ti 3}{\ti P}_{\ti 1 \ti 2})\, \bS_{2\ti 2}\, \bS_{3\ti 3}
 \Big\}=0\,,
 }
 \end{array}
 \eq
i.e.
 \beq\label{e52}
  \begin{array}{c}
  \displaystyle{
 \tr_{2,3,\ti2,\ti 3} \Big\{ \p_\hbar R_{13}^\hbar\,  [{\ti P}_{\ti 1 \ti 2},{\ti
 P}_{\ti 1 \ti 3}]\, \bS_{2\ti 2}\, \bS_{3\ti 3}
 \Big\}=
 \tr_{2,3} \Big\{ \p_\hbar R_{13}^\hbar\,   [\bS_{2\ti 1}\,, \bS_{3\ti
 1}]
 \Big\}=0
 }
 \end{array}
 \eq
and, therefore,
 \beq\label{e53}
  \begin{array}{c}
  \displaystyle{
 tr_{2}\,\bS_{2\ti 1}\sim 1_M\,,
 }
 \end{array}
 \eq
that is {\em matrix analogue of the variable $\tr S$ (or $S_0$ in
basis $\{T_\al\}$) should be not an arbitrary $\Matm$ matrix but the
one proportional to identity matrix $1_M$}. It is easy to see that
the coefficient behind $1_M$ in (\ref{e53}) should be a constant on
equations of motion (since it equals $tr_{2\ti 1}\,\bS_{2\ti 1}/M$)
Therefore, the next set of constraints is generated by
 \beq\label{e54}
  \begin{array}{c}
  \displaystyle{
 tr_{1}\,{\dot \bS}_{1\ti 1}=0
 }
 \end{array}
 \eq
which means that (\ref{e53}) should be preserved by dynamics of
equations of motion.

\paragraph{Equations of motion and Lax pairs.} On constraints (\ref{e53}), when (\ref{e51}) is true, we have the
following equation obtained by subtracting (\ref{e50}) from
(\ref{e49}):
 \beq\label{e55}
  \begin{array}{c}
  \displaystyle{
 \tr_{2,3,\ti2,\ti 3} \Big\{ \left([R^\hbar_{12,\ti 1\ti 2},
 R^{\hbar}_{23,\ti 2\ti 3}]-[R^{\hbar}_{13,\ti 1\ti 3},r_{12,\ti 1\ti 2}]-[r_{23,\ti 2\ti 3},R^\hbar_{13,\ti 1\ti
 3}]\right)\,
  \bS_{2\ti 2}\, \bS_{3\ti 3}
 \Big\}=0\,,
 }
 \end{array}
 \eq
It is analogous to $\tr_{2,3}\{(eq.\ (\ref{e25}))S_2S_3\}$, which
underlied the Lax equations in scalar case. For a similar reason we
obtain the following equations of motion in relativistic case:
 \beq\label{e56}
  \begin{array}{c}
  \displaystyle{
 {\dot \bS_{1\ti 1}}=[\bS_{1\ti 1},J_1^\eta(\bS_{1\ti 1})]\,,
 }
 \end{array}
 \eq
where
 \beq\label{e57}
  \begin{array}{c}
  \displaystyle{
 J_1^\eta(\bS_{1\ti 1})=\tr_{2\ti 2}( (R_{12,\ti 1\ti 2}^{\eta,(0)}-r_{12,\ti 1\ti 2}^{(0)})\,\bS_{2\ti 2}
 )=\tr_{2\ti 2}( (R_{12}^{\eta,(0)}-r_{12}^{(0)})\otimes{\ti P}_{\ti 1\ti 2}\,\bS_{2\ti 2} )=
 }
 \\ \ \\
  \displaystyle{
 =\tr_{2}( (R_{12}^{\eta,(0)}-r_{12}^{(0)})\,\bS_{2\ti 1} )\,.
 }
 \end{array}
 \eq
In scalar case $M=1$ the latter equation turns into (\ref{e22}). The
Lax pair is given by
  \beq\label{e58}
  \begin{array}{c}
  \displaystyle{
L^\eta(z,\bS)=\tr_{2,\ti 2}\left( R^{\,\eta}_{12,\ti 1\ti 2}(z)\,
\bS_{2\ti 2} \right)\,, \quad M^\eta(z,\bS)=\tr_{2,\ti 2}\left(
r_{12,\ti 1\ti 2}(z)\, \bS_{2\ti 2} \right)\,.
 }
 \end{array}
 \eq
In non-relativistic case equations of motion are
 \beq\label{e59}
  \begin{array}{c}
  \displaystyle{
 {\dot \bS_{1\ti 1}}=[\bS_{1\ti 1},J_1(\bS_{1\ti 1})]\,,
 }
 \end{array}
 \eq
where
 \beq\label{e60}
  \begin{array}{c}
  \displaystyle{
 J_1(\bS_{1\ti 1})=\tr_{2\ti 2}( m_{12,\ti 1\ti 2}(0)\,\bS_{2\ti 2}
 )=\tr_{2\ti 2}( m_{12}(0)\otimes{\ti P}_{\ti 1\ti 2}\,\bS_{2\ti 2} )=
 \tr_{2}( m_{12}(0)\,\bS_{2\ti 1} )\,.
 }
 \end{array}
 \eq
The Lax pair is given by
  \beq\label{e61}
  \begin{array}{c}
  \displaystyle{
L(z,\bS)=\tr_{2,\ti 2}\left( r_{12,\ti 1\ti 2}(z)\, \bS_{2\ti 2}
\right)\,, \quad M(z,\bS)=\tr_{2,\ti 2}\left( m_{12,\ti 1\ti 2}(z)\,
\bS_{2\ti 2} \right)\,.
 }
 \end{array}
 \eq
Let us stress again that together with (\ref{e56}) or (\ref{e59})
the constraints (\ref{e53}), (\ref{e54}) should be fulfilled. Below
we will see that these constraints are fulfilled for a special class
of elliptic matrix tops.

\paragraph{Matrix generalization of $\mZ_2$ reduced elliptic tops.}
We start with non-relativistic case.
 Similarly to (\ref{e041}) and following (\ref{e61}) we have the
 following Lax pair for the matrix elliptic top:
 \beq\label{e63}
 \begin{array}{c}
  \displaystyle{
L(z,\bS)=\sum\limits_{\al\neq 0}T_\al\otimes
\bS_\al\,\vf_\al(z,\om_\al)\,,\quad M(z,\bS)=\sum\limits_{\al\neq
0}T_\al\otimes \bS_\al\, f_\al(z,\om_\al)\,.
 }
 \end{array}
 \eq
The r.h.s. of the Lax equation
 \beq\label{e64}
 \begin{array}{c}
  \displaystyle{
\frac{d}{dt}\,L(z,\bS)=[L(z,\bS),M(z,\bS)]
 }
 \end{array}
 \eq
is equal to
 \beq\label{e65}
 \begin{array}{c}
  \displaystyle{
[L(z,\bS),M(z,\bS)]=\sum\limits_{\be,\ga\neq 0}
 T_\be T_\ga\otimes \bS_\be\bS_\ga\,\vf_\be(z)f_\ga(z)-T_\be T_\ga\otimes
 \bS_\be\bS_\ga\,f_\be(z)\vf_\ga(z)=
 }
 \end{array}
 \eq
By symmetrizing indices $\be$ and $\ga$ we get (here for short we
use $\vf_\be(z)=\vf_\be(z,\om_\be)$ and the same for $f_\be(z)$)
 \beq\label{e66}
 \begin{array}{c}
  \displaystyle{
=\sum\limits_{\be,\ga\neq 0}
 \frac12\, T_\be T_\ga\otimes \bS_\be\bS_\ga\,\vf_\be(z)f_\ga(z)
 +\frac12\, T_\ga T_\be\otimes \bS_\ga\bS_\be\,\vf_\ga(z)f_\be(z)
 }
 \\ \ \\
  \displaystyle{
\ \ \ \ \ \ \ -\frac12\, T_\be T_\ga\otimes
\bS_\be\bS_\ga\,f_\be(z)\vf_\ga(z)
 -\frac12\, T_\ga T_\be\otimes \bS_\ga\bS_\be\,f_\ga(z)\vf_\be(z)
 =
 }
 \\ \ \\
  \displaystyle{
=\sum\limits_{\be,\ga\neq 0} \frac12\, \Big( T_\be T_\ga\otimes
\bS_\be\bS_\ga - T_\ga T_\be\otimes \bS_\ga\bS_\be \Big) \Big(
\vf_\be(z)f_\ga(z)-\vf_\ga(z)f_\be(z)
\Big)\stackrel{(\ref{e9115})}{=}
 }
 \end{array}
 \eq
 \beq\label{e67}
 \begin{array}{c}
  \displaystyle{
=\sum\limits_{\be,\ga\neq 0} \frac12\,T_{\be+\ga}\otimes \Big(
 \kappa_{\be,\ga}\,\bS_\be\bS_\ga - \kappa_{\ga,\be}\,\bS_\ga\bS_\be \Big) \vf_{\be+\ga}(z) \Big(
 E_2(\om_\be)-E_2(\om_\ga)
\Big)=
 }
 \\ \ \\
  \displaystyle{
=\sum\limits_{\be,\ga\neq 0} T_{\be+\ga}\otimes \Big(
 \kappa_{\be,\ga}\,\bS_\be\bS_\ga - \kappa_{\ga,\be}\,\bS_\ga\bS_\be \Big)\,
 \vf_{\be+\ga}(z)\,
 J_\ga\,,
 }
 \end{array}
 \eq
where $J_\ga=-E_2(\om_\ga)$ as in (\ref{e0400}) and
$\kappa_{\ga,\be}$ are structure constants (\ref{e905}). Finally,
equations of motion take the form
 \beq\label{e68}
 \begin{array}{c}
  \displaystyle{
{\dot \bS}_\al=\sum\limits_{\be,\ga:\,\be+\ga=\al} \Big(
 \kappa_{\be,\ga}\,\bS_\be\bS_\ga - \kappa_{\ga,\be}\,\bS_\ga\bS_\be \Big)\,
 J_\ga\,,\ \al\neq 0;\ J_\ga=-E_2(\om_\ga)\,.
 }
 \end{array}
 \eq
In scalar case $M=1$ the latter equations coincide with
(\ref{e0411}).

In the above equations we did not include $\al=0$ component into the
Lax pair (\ref{e63}), i.e. $\bS_0=0$, and therefore (\ref{e53}) is
fulfilled. However (\ref{e54}) is not fulfilled. Indeed, for
$\al=\be+\ga=0$ in (\ref{e66}) we need to use (\ref{e9125}) instead
of (\ref{e9115}). It yields ($\kappa_{\be,-\be}=1$) the following
explicit expression for (\ref{e54}):
 \beq\label{e69}
 \begin{array}{c}
  \displaystyle{
0={\dot \bS}_0=\sum\limits_{\be\neq 0}\,
 [\bS_\be,\bS_{-\be}]\,
 E_2'(\om_\be)\,.
 }
 \end{array}
 \eq
It is nontrivial because $E_2'(z)$ is an odd function. A natural way
to fulfill this constraint is to set
 \beq\label{e70}
 \begin{array}{c}
  \displaystyle{
\chi:\ \ \bS_\al=\bS_{-\al}\,,\ \ \hbox{for}\ \hbox{all}\ \al\,.
 }
 \end{array}
 \eq
It is matrix analogue of $\mZ_2$ reduced elliptic top defined by
(\ref{e0855}).

The set of constraints (\ref{e70}) is preserved by dynamics
(\ref{e68}):
 \beq\label{e71}
 \begin{array}{c}
  \displaystyle{
{\dot \bS_{-\al}}\left.\right|_\chi={\dot
\bS_{\al}}\left.\right|_\chi
 }
 \end{array}
 \eq
since $\kappa_{-\be,-\ga}=\kappa_{\be,\ga}$ and $J_\al=J_{-\al}$.
Therefore, we have well defined matrix valued elliptic top given by
the Lax pair (\ref{e63}), equations of motion (\ref{e68}) and
$\mZ_2$ reduction constraints (\ref{e70}).

In relativistic case we have the following direct generalization of
(\ref{e042}):
 \beq\label{e72}
 \begin{array}{c}
  \displaystyle{
{L^\eta}(z,\bS)=\sum\limits_{\al} T_\al\otimes \bS_\al\,
\vf_\al(z,\om_\al+\eta)\,,\quad M^\eta(z,\bS)=-\sum\limits_{\al\neq
0}T_\al\otimes \bS_\al\,\vf_\al(z,\om_\al)\,.
 }
 \end{array}
 \eq
The Lax equations lead to equations of motion (\ref{e59}) for
matrix-variables
 \beq\label{e73}
 \begin{array}{c}
  \displaystyle{
{\dot \bS}_\al=\sum\limits_{\be,\ga:\,\be+\ga=\al} \Big(
 \kappa_{\be,\ga}\,\bS_\be\bS_\ga - \kappa_{\ga,\be}\,\bS_\ga\bS_\be \Big)\,
 J_\ga^\eta\,,\ \al\neq 0;\ \ J_\al^\eta=E_1(\eta+\om_\al)-E_1(\om_\al)
 }
 \end{array}
 \eq
via (\ref{e911}). The constraints (\ref{e53}), (\ref{e54}) means
that
 \beq\label{e74}
 \begin{array}{c}
  \displaystyle{
{ \bS}_0=S_0\, 1_M\,,
 }
 \end{array}
 \eq
where $1_M$ is identity $M\times M$ matrix, and
  \beq\label{e75}
  \begin{array}{c}
    \displaystyle{
\frac{1}{\vf_\al(\eta,\om_\al)}\,\bS_\al=\frac{1}{\vf_{-\al}(\eta,-\om_\al)}\,\bS_{-\al}\,,
\ \al\neq 0\,.
 }
 \end{array}
 \eq

{\em Let us now mention that the derivation of equations of motion
from the Lax pairs (\ref{e63}) or (\ref{e72}) did not use that $\bS$
is a matrix. In fact, we can perform the same calculation thinking
of $\bS_\al$ as elements of associative and noncommutative algebra.}



\section{Noncommutative Painlev\'e VI equation}
As it was explained in the Introduction the Lax pair (\ref{e041}) of
the non-relativistic top (\ref{e0411}) satisfies also the monodromy
preserving condition  (\ref{e0870}) and provides in this way the
non-autonomous version of the Euler-Arnold equations (\ref{e0414}).
This construction is straightforwardly generalized to the matrix
extension of elliptic top described by the Lax pair (\ref{e63}).
Namely, we have the following statement.
 \begin{predl}
 The Lax pair
 $$
  \displaystyle{
L(z,\bS)=\sum\limits_{\al\neq 0}T_\al\otimes
\bS_\al\,\vf_\al(z,\om_\al)\,,\quad M(z,\bS)=\sum\limits_{\al\neq
0}T_\al\otimes \bS_\al\, f_\al(z,\om_\al)\,. }
 $$
 with $\mZ_2$ reduction condition $\bS_\al=\bS_{-\al}$
satisfies the monodromy preserving condition
 $$
  \displaystyle{
\frac{d}{d\tau}\,L(w,\bS)-\frac{1}{2\pi\imath}\frac{\p}{\p
w}\,M(w,\bS)=[L(w,\bS),M(w,\bS)]
 }
 $$
 and provides non-autonomous version of the matrix top equations:
 \beq\label{e595}
  \begin{array}{c}
  \displaystyle{
 \frac{d}{d\tau}\,{\bS_{1\ti 1}}=[\bS_{1\ti 1},J_1(\bS_{1\ti 1})]\,,
 }
 \end{array}
 \eq
or
 \beq\label{e685}
 \begin{array}{c}
  \displaystyle{
 \frac{d}{d\tau}\,{\bS}_\al=\sum\limits_{\be,\ga:\,\be+\ga=\al} \Big(
 \kappa_{\be,\ga}\,\bS_\be\bS_\ga - \kappa_{\ga,\be}\,\bS_\ga\bS_\be \Big)\,
 J_\ga\,,\ \al\neq 0;\ J_\ga=-E_2(\om_\ga)\,.
 }
 \end{array}
 \eq
 \end{predl}
As in the scalar case the proof is based on the heat equation
$2\pi\imath\,\p_\tau\vf_\al(z,\om_\al)=\p_z f_\al(z,\om_\al)$.

In the same way one can define matrix extension of the
non-autonomous version of the Zhukovsky-Volterra
gyrostat\footnote{The autonomous version is of course well defined
also. One should just replace the $\tau$-derivative by
$t$-derivative.} in $N=2$ case. The constants $\nu'_1,\nu'_2,\nu'_3$
are kept to be scalar, i.e.
 \beq\label{e08784}
 \begin{array}{c}
  \displaystyle{
[\bS_\al,\nu'_\be]=0\,.
 }
 \end{array}
 \eq
 \begin{predl}
 The Lax pair from $\hbox{Mat}(2,\mC)\otimes \Matm$
 \beq\label{e08785}
 \begin{array}{c}
  \displaystyle{
L^{PVI}(w,\bS)=L^{ZV}(w)\stackrel{(\ref{e0875})}{=}\frac{1}{2\imath}\sum\limits_{\al=1}^3\sigma_\al\otimes
\left(\bS_\al\vf_\al(w,\om_\al)+1_{M\times
M}{{\tilde\nu}_\al}{\vf_\al(w-\om_\al,\om_\al)}\right)\,,}
\\ \ \\
  \displaystyle{
M^{PVI}(w,\bS)=-\frac{1}{2\imath}\sum\limits_{\al=1}^3
\sigma_\al\otimes \bS_\al
\frac{\vf_1(w,\om_1)\vf_2(w,\om_2)\vf_3(w,\om_3)}{\vf_\al(w,\om_\al)}
+E_1(w)L^{PVI}(w,\bS)\,.
 }
 \end{array}
 \eq
provides through substitution into the monodromy preserving
condition (\ref{e0870}) the following equations
 \beq\label{e08786}
 \begin{array}{c}
  \displaystyle{
\frac{d}{d\tau}\,\bS_\al=\frac{1}{2}\,(\bS_\be\bS_\ga+\bS_\ga\bS_\be)(E_2(\om_\be)-E_2(\om_\ga))+\bS_\be\nu'_\ga-\bS_\ga\nu'_\be\,,
 }
 \end{array}
 \eq
 $$
 \omega_1=\tau/2\,,~\omega_2=(1+\tau)/2\,,~\omega_3=1/2\,.
 $$
where $(\al,\be,\ga)=(1,2,3)$ up to cyclic permutations. In matrix
form we have
 \beq\label{e08787}
  \begin{array}{c}
  \displaystyle{
 \frac{d}{d\tau}\,{\bS_{1\ti 1}}=[\bS_{1\ti 1},J_1(\bS_{1\ti 1})+\nu'\otimes 1_{M\times
 M}]\,.
 }
 \end{array}
 \eq
 \end{predl}
 In the scalar ($N=1)$ case equation (\ref{e08786}) or
 (\ref{e08787}) turns into the non-autonomous Zhukovsky-Volterra
 gyrostat (\ref{e0111}), which is known to be equivalent to
 Painlev\'e VI equation. By this reason we call (\ref{e08786}) or
 (\ref{e08787}) noncommutative Painlev\'e VI equation. Here we
 should repeat the Remark 1 from the end of Introduction that equations (\ref{e08786}) keep the same form if $\bS$ take values
in an arbitrary non-commutative associative algebra $\mathcal{A}$.

\section{Special elliptic Gaudin models as matrix tops}
\setcounter{equation}{0}

Consider the following ${\rm gl}_N$ Lax pair given by $N\times N$
matrices
 \beq\label{e76}
 \begin{array}{c}
  \displaystyle{
L^G(z)=A^0+\sum\limits_{\al\neq 0} A^\al \vf_\al(z,\om_\al)\,,\quad
M^G(z)=\sum\limits_{\al\neq 0}A^\al f_\al(z,\om_\al)\,,
 }
 \end{array}
 \eq
where $A^\al\in\Mat$ is a set of ${\rm gl}_N$-valued matrices with
constraints
 \beq\label{e760}
 \begin{array}{c}
  \displaystyle{
A^0=1_N\, S_0\,,
 }
 \end{array}
 \eq
 \beq\label{e761}
 \begin{array}{c}
  \displaystyle{
A^\al=A^{-\al}\ \ \hbox{for all}\ \al\neq 0\,,
 }
 \end{array}
 \eq
which are similar to (\ref{e70}).
It can be viewed as a special elliptic Gaudin model. Indeed, it
follows from quasiperiodic properties
  \beq\label{e77}
  \begin{array}{c}
    \displaystyle{
\phi(z+1,u)=\phi(z,u)\,,\quad \phi(z+\tau,u)=\exp(-2\pi\imath
u)\phi(z,u)
 }
 \end{array}
 \eq
that (for $\al\neq 0$)
  \beq\label{e78}
  \begin{array}{c}
    \displaystyle{
 \vf_\al(z+1,\om_\al)=\exp(2\pi\imath\frac{\al_2}{N})\,\vf_\al(z,\om_\al)\,,
 }
 \\ \ \\
    \displaystyle{
 \vf_\al(z+\tau,\om_\al)=\exp(-2\pi\imath\frac{\al_1}{N})\,\vf_\al(z,\om_\al)\,.
 }
 \end{array}
 \eq
Therefore, functions (sections of bundle) $\{\vf_\al(z)\}$ are
double-periodic on a "large" torus $\Sigma_{N,N\tau}$ generated by
fundamental parallelogram with periods $N,N\tau$. The latter means
that $L^G(z)$ is a double-periodic function on $\Sigma_{N,N\tau}$
with $N^2-1$ simple poles at points $N\om_\al=\al_1+\al_2\tau$,
$\al\neq 0$. The residues at these points are linear combinations of
$A^\be$:
  \beq\label{e79}
  \begin{array}{c}
    \displaystyle{
\res\limits_{z=N\om_\al} L^G(z)=\sum\limits_{\be\neq 0}
\kappa_{\be,\al}^2 A^\be\,,
 }
 \end{array}
 \eq
where $\kappa_{\be,\al}$ is given by (\ref{e905}). This is why we
refer to this model as Gaudin one. The Lax equations are equivalent
to
  \beq\label{e80}
  \begin{array}{c}
    \displaystyle{
{\dot A}^\al=\sum\limits_{\be+\ga=\al}[A^\be,A^\ga]\, J_\ga\,,\quad
J_\ga=-E_2(\om_\ga)\,,\ \al\neq 0\,.
 }
 \end{array}
 \eq
These equations generalize the elliptic top equations of motion
(\ref{e0411}) in the following sense. Equations (\ref{e0411}) are
reproduced from (\ref{e80}) via reduction
  \beq\label{e81}
  \begin{array}{c}
    \displaystyle{
A^\al=T_\al S_\al\,.
 }
 \end{array}
 \eq
At the same time (\ref{e761}) reduces to (\ref{e0855}), i.e.
(\ref{e80}) can  be viewed as matrix generalization of $\mZ_2$
reduced elliptic top.

 As in (\ref{e69}) the constraints (\ref{e761}) fulfill the
 constrain
  \beq\label{e82}
  \begin{array}{c}
    \displaystyle{
0={\dot A}^0=\sum\limits_{\be}\,[A^\be,A^{-\be}]\, E_2'(\om_\be)\,,
 }
 \end{array}
 \eq
which appear from "zero mode" of the Lax equations. In the same way,
similarly to (\ref{e71}) ${\dot A^\al}={\dot A^{-\al}}$ on
constraints (\ref{e761}), i.e. these constraints are preserved by
dynamics.

Similarly to results of the previous section we can easily construct
non-autonomous models generalizing (\ref{e80}) through the monodromy
preserving condition (\ref{e0870}). The answer is as follows:
  \beq\label{e84}
  \begin{array}{c}
    \displaystyle{
\frac{d}{d\tau}\,{A}^\al=\sum\limits_{\be+\ga=\al}[A^\be,A^\ga]\,
J_\ga\,,\quad J_\ga=-E_2(\om_\ga)\,,\ \al\neq 0\,.
 }
 \end{array}
 \eq
It is interesting to mention that in $N=2$ case these equations are
equivalent to the Painlev\'e VI equation (\ref{e0874}) after
reduction by coadjoint action of "common" ${\rm GL}(2,\mC)$:
$A^\al\rightarrow gA^\al g^{-1}$. See details in \cite{Zot}.







\section{Appendix: elliptic functions and $R$-matrices}
\def\theequation{A.\arabic{equation}}
\setcounter{equation}{0}

The Baxter-Belavin $R$-matrix as well as elliptic tops uses special
basis in $\Mat$. Let
 \beq\label{e903}
 \begin{array}{c}
  \displaystyle{
Q_{kl}=\delta_{kl}\exp(\frac{2\pi
 \imath}{N}k)\,,\ \ \ \Lambda_{kl}=\delta_{k-l+1=0\,{\hbox{\tiny{mod}}}
 N}\,,\quad Q^N=\Lambda^N=1_{N\times N}\,.
 }
 \end{array}
 \eq
 Then for
 \beq\label{e904}
 \begin{array}{c}
  \displaystyle{
 T_a=T_{a_1 a_2}=\exp\left(\frac{\pi\imath}{N}\,a_1
 a_2\right)Q^{a_1}\Lambda^{a_2}\,,\quad
 a=(a_1,a_2)\in\mZ_N\times\mZ_N
 }
 \end{array}
 \eq
due to
 \beq\label{e9041}
 \begin{array}{c}
  \displaystyle{
 \exp\left(\frac{2\pi\imath}{N}\,a_1
 a_2\right)Q^{a_1}\Lambda^{a_2}=\Lambda^{a_2}Q^{a_1}
 }
 \end{array}
 \eq
we have
  \beq\label{e905}
 \begin{array}{c}
  \displaystyle{
T_\al T_\be=\kappa_{\al,\be} T_{\al+\be}\,,\ \ \
\kappa_{\al,\be}=\exp\left(\frac{\pi \imath}{N}(\be_1
\al_2-\be_2\al_1)\right)\,,
 }
 \end{array}
 \eq
where $\al+\be=(\al_1+\be_1,\al_2+\be_2)$. The structure constant
$\ka_{\al,\be}$ satisfy
  \beq\label{e9051}
 \begin{array}{c}
  \displaystyle{
\sum\limits_{\al} \ka_{\al,\ga}^2=N^2\,\delta_{\ga,0}\,,
 }
 \end{array}
 \eq
which is equivalent to identity $P_{12}^2=1\otimes 1$ for the
permutation operator $P_{12}$ given by
  \beq\label{e9052}
 \begin{array}{c}
  \displaystyle{
P_{12}=\frac{1}{N}\sum\limits_{\al\in\,\mZ_N\times\mZ_N} T_\al
\otimes T_{-\al}=\sum\limits_{i,j=1}^N E_{ij}\otimes E_{ji}\,.
 }
 \end{array}
 \eq
From (\ref{e905}) we obviously get
  \beq\label{e906}
 \begin{array}{c}
  \displaystyle{
[T_\al, T_\be]=C_{\al,\be} T_{\al+\be}\,,\ \ \
C_{\al,\be}=\kappa_{\al,\be}-\kappa_{\be,\al}\,,
 }
 \end{array}
 \eq
i.e. the set $\{T_\al\}$ can be also considered as a basis in ${\rm
gl}_N$ Lie algebra.  It is also called the sin-algebra basis since
$C_{\al,\be}=2\imath \sin(\frac{\be_1 \al_2-\be_2\al_1}{2N})$. Being
written in such a form it has natural generalization to ${\rm
gl}_\infty$. From the point of view of integrable systems it
corresponds to (Arnold's type) 2D hydrodynamics.

For $N=2$ we have
 $$
 Q=\mats{-1}{0}{0}{1}\quad \quad \Lambda=\mats{0}{1}{1}{0}\,,
 $$
and, therefore, $\{T_\al\}$ in this case is the set of Pauli
matrices:
  \beq\label{e9061}
 \begin{array}{c}
  \displaystyle{
T_{00}=\si_0=1_{2\times 2}\,,\ \ T_{10}=-\si_3\,,\ \
T_{01}=\si_1\,,\ \ T_{11}=\si_2\,.
 }
 \end{array}
 \eq

 \subsection{Elliptic functions}

\noindent {\bf The Kronecker and Eisenstein functions \cite{Weil}.}
The following set of elliptic functions\footnote{To be exact, some
of these function are not double-periodic. In this sense they are
not functions but rather sections of bundles (the Kronecker
functions) or components of connections ($E_1$-function). See the
quasi-periodic properties e.g. in \cite{LOZ10}.} on elliptic curve
$\mC/\mZ\oplus\tau\mZ$ with moduli $\tau$ (Im$\tau>0$) is widely
used in this paper:

The Kronecker function
  \beq\label{e907}
  \begin{array}{l}
  \displaystyle{
 \phi(\eta,z)=\frac{\vth'(0)\vth(\eta+z)}{\vth(\eta)\vth(z)}
 }
 \end{array}
 \eq
is defined in terms of the odd Riemann theta-function
 \beq\label{e9077}
 \begin{array}{c}
  \displaystyle{
\vth(z)=\vth(z|\tau)=\displaystyle{\sum _{k\in \mathbb Z}} \exp
\left ( \pi \imath \tau (k+\frac{1}{2})^2 +2\pi \imath
(z+\frac{1}{2})(k+\frac{1}{2})\right )\,.
  }
 \end{array}
 \eq
In rational and trigonometric cases it equals $1/\eta+1/z$ and
$\coth(\eta)+\coth(z)$ respectively. The derivative of the Kronecker
function
  \beq\label{e9071}
  \begin{array}{l}
  \displaystyle{
 f(z,u)\equiv\p_u\phi(z,u)=\phi(z,u)(E_1(z+u)-E_1(u))
 }
 \end{array}
 \eq
uses the definition of the first Eisenstein function:
  \beq\label{e908}
  \begin{array}{c}
  \displaystyle{
 E_1(z)=\vth'(z)/\vth(z)\,.
 }
 \end{array}
 \eq
It is odd. In rational and trigonometric cases it equals $1/z$ and
$\coth(z)$ respectively. Its derivative
   \beq\label{e9081}
  \begin{array}{c}
  \displaystyle{
   E_2(z)= -\p_z
   E_1(z)=\wp(z)-\frac{1}{3}\frac{\vth'''(0)}{\vth'(0)}\,,\quad
   \Big(\,\hbox{and}\ E_1(z)=\zeta(z)+\frac{z}{3}\frac{\vth'''(0)}{\vth'(0)}\,\Big)
 }
 \end{array}
 \eq
 is known as the second Eisenstein function. The functions $\wp(z)$ and $\zeta(z)$ are the
 Weierstrass $\wp$- and $\zeta$-functions.

The local expansion of the Kronecker and Eisenstein functions near
$z=0$:
 \beq\label{e9082}
 \begin{array}{c}
  \displaystyle{
 \phi(z,u)=\frac{1}{z}+E_1(u)+\frac{z}{2}\,(E_1^2(u)-\wp(u))+
 O(z^2)\,,
  }
 \end{array}
 \eq
 \beq\label{e9083}
 \begin{array}{c}
  \displaystyle{
 E_1(z)=\frac{1}{z}+\frac{z}{3}\,\frac{\vth'''(0)}{\vth'(0)}+O(z^3)\,.
  }
 \end{array}
 \eq
 In particular, we conclude from (\ref{e9082}) that
  \beq\label{e9084}
 \begin{array}{c}
  \displaystyle{
 f(0,u)=-E_2(u)\,.
  }
 \end{array}
 \eq
  The Kronecker function satisfies the heat equation
 \beq\label{A.4b}
\p_\tau\phi(u,w)-\frac{1}{2\pi i}\p_u\p_w\phi(u,w)=0\,.
 \eq

Most of the Lax equations are due to the Fay trisecant identity
  \beq\label{e909}
  \begin{array}{c}
  \displaystyle{
\phi(z,q)\phi(w,u)=\phi(z-w,q)\phi(w,q+u)+\phi(w-z,u)\phi(z,q+u)
 }
 \end{array}
 \eq
 and its degenerations
  \beq\label{e911}
  \begin{array}{c}
  \displaystyle{
 \phi(z,q)\phi(w,q)=\phi(z+w,q)(E_1(z)+E_1(w)+E_1(q)-E_1(z+w+q))\,.
 }
 \end{array}
 \eq
  \beq\label{e9115}
  \begin{array}{c}
  \displaystyle{
 \phi(z,x)f(z,y)-\phi(z,y)f(z,x)=\phi(z,x+y)(E_2(x)-E_2(y))\,,
 }
 \end{array}
 \eq
  \beq\label{e912}
  \begin{array}{c}
  \displaystyle{
 \phi(\hbar,z)\phi(\hbar,-z)=\wp(\hbar)-\wp(z)=E_2(\hbar)-E_2(z)\,,
 }
 \end{array}
 \eq
  \beq\label{e9125}
  \begin{array}{c}
  \displaystyle{
 \phi(z,x)f(z,-x)-\phi(z,-x)f(z,x)=E_2'(x)=\wp'(x)\,.
 }
 \end{array}
 \eq
The definition (\ref{e050}) of the Baxter-Belavin $R$-matrix uses
the set of $N^2$ functions
 \beq\label{e910}
 \begin{array}{c}
  \displaystyle{
 \vf_\al^\hbar(z)\equiv\vf_\al(z,\om_\al+\hbar)=\exp(2\pi\imath\,z\,\p_\tau\om_\al)\,\phi(z,\hbar+\om_\al)\,,
 }
 \end{array}
 \eq
where
 \beq\label{e9101}
 \begin{array}{c}
  \displaystyle{
\om_\al=\frac{\al_1+\al_2\tau}{N}\,,\quad
 \p_\tau\om_\al=\frac{\al_2}{N}\,,\quad
 \al=(\al_1,\al_2)\in \mZ_N\times\mZ_N\,,
 }
 \end{array}
 \eq
 The following notations are also used for $\al\neq 0$ (i.e. $(\al_1,\al_2)\neq
 (0,0)$):
 \beq\label{e9102}
 \begin{array}{c}
  \displaystyle{
 \vf_\al(z,\om_\al)=\vf_\al^0(z)=\exp(2\pi\imath\,z\,\p_\tau\om_\al)\,\phi(z,\om_\al)\,,
 }
 \end{array}
 \eq
 \beq\label{e9103}
 \begin{array}{c}
  \displaystyle{
 f_\al(z,\om_\al)=\exp(2\pi\imath\,z\,\p_\tau\om_\al)\,f(z,\om_\al)\,.
 }
 \end{array}
 \eq
The index $\al$ in $\vf_\al$ and $f_\al$ reminds about the
exponential factor.

\subsection{$R$-matrix structures for elliptic tops}
\def\theequation{B.\arabic{equation}}
\setcounter{equation}{0}

Let us list explicit formulae for the coefficients of expansions
(\ref{e053})-(\ref{e055}). First, write down again the
Baxter-Belavin $R$-matrix (\ref{e050}) with both arguments in
$\vf$-functions (see notations (\ref{e910})-(\ref{e9103})):
 \beq\label{e921}
 \begin{array}{c}
  \displaystyle{
 R^{\,\hbar}_{12}(z)=
 \sum\limits_{\al\in\,{\mathbb Z}_N\times{\mathbb Z}_N} \vf_\al(z,\om_\al+\hbar)\, T_\al\otimes
 T_{-\al}\,.
 }
 \end{array}
 \eq
Using (\ref{e9082}), (\ref{e9083})  we obtain the classical
Belavin-Drinfeld $r$-matrix
 \beq\label{e922}
 \begin{array}{c}
  \displaystyle{
 r_{12}(z)=E_1(z)\, 1\otimes 1 +
 \sum\limits_{\al\neq 0} \vf_\al(z,\om_\al)\, T_\al\otimes
 T_{-\al}\,.
 }
 \end{array}
 \eq
It satisfies the classical Yang-Baxter equation
  \beq\label{e057}
  \begin{array}{c}
  \displaystyle{
[r_{12},r_{13}]+[r_{12},r_{23}]+[r_{13},r_{23}]=0\,,\quad
r_{ab}=r_{ab}(z_a-z_b)
 }
 \end{array}
 \eq
due to the quantum one (\ref{e051}). The next
 term in (\ref{e053}):
 \beq\label{e923}
 \begin{array}{c}
  \displaystyle{
 m_{12}(z)=\frac{1}{2}\,(E_1^2(z)-\wp(z))\, 1\otimes 1 +
 \sum\limits_{\al\neq 0} f_\al(z,\om_\al)\, T_\al\otimes
 T_{-\al}=
 }
 \\
  \displaystyle{
 =\frac12\Big(r_{12}^2(z)-1\otimes 1\, N^2\wp(z)\Big)\,.
 }
 \end{array}
 \eq
The second line follows from the unitarity condition (\ref{e052}).

Using local expansion (\ref{e9082}) we obtain the terms from
(\ref{e054}), (\ref{e055}):
 \beq\label{e924}
 \begin{array}{c}
  \displaystyle{
 R_{12}^{\hbar,(0)}=
 \sum\limits_{\al} (E_1(\hbar+\om_\al)+2\pi\imath\p_\tau\om_\al)\, T_\al\otimes
 T_{-\al}\,,
 }
 \end{array}
 \eq
 \beq\label{e925}
 \begin{array}{c}
  \displaystyle{
 r_{12}^{(0)}=
 \sum\limits_{\al\neq 0} (E_1(\om_\al)+2\pi\imath\p_\tau\om_\al)\, T_\al\otimes
 T_{-\al}\,.
 }
 \end{array}
 \eq

\vskip2mm

\noindent {\bf Properties and identities.} The skew-symmetry
(\ref{e11}) of the quantum $R$ matrix (\ref{e921}) as well as the
unitarity (\ref{e052}) leads to
 \beq\label{e926}
 \begin{array}{c}
  \displaystyle{
  r_{21}(-z)\equiv P_{12}\,r_{12}(-z)\,P_{12}=-r_{12}(z)\,,\ \ \
 m_{12}(z)=m_{21}(-z)\,,
  }
 \end{array}
 \eq
 \beq\label{e927}
 \begin{array}{c}
  \displaystyle{
 R^{\hbar,(0)}_{12}=-R^{-\hbar,(0)}_{21}\,,\quad r_{12}^{(0)}=-r_{21}^{(0)}\,.
  }
 \end{array}
 \eq

Various formulae relating the coefficients follow from the
associative Yang-Baxter equation (\ref{e10}) (and the original
Yang-Baxter equation (\ref{e051})). In particular, in the limit
$\eta\rightarrow\hbar$ it gives:
 \beq\label{e931}
  \begin{array}{c}
  \displaystyle{
 R^\hbar_{12}
 R^{\hbar}_{23}=R^{\hbar}_{13}r_{12}+r_{23}R^\hbar_{13}-\p_\hbar
 R_{13}^\hbar\,.
 }
 \end{array}
 \eq
By changing indices $1\leftrightarrow3$ (i.e. conjugating equation
by $P_{13}$ and renaming $z_1\leftrightarrow z_3$), changing also
$\hbar\rightarrow -\hbar$ and then using skew-symmetry (\ref{e11})
it transforms into
 \beq\label{e932}
  \begin{array}{c}
  \displaystyle{
 R^\hbar_{23}
 R^{\hbar}_{12}=R^{\hbar}_{13}r_{23}+r_{12}R^\hbar_{13}-\p_\hbar
 R_{13}^\hbar\,.
 }
 \end{array}
 \eq
Subtracting (\ref{e932}) from (\ref{e931}) yields
 \beq\label{e933}
  \begin{array}{c}
  \displaystyle{
 [R^\hbar_{12},
 R^{\hbar}_{23}]=[R^{\hbar}_{13},r_{12}]-[R^{\hbar}_{13},r_{23}]\,.
 }
 \end{array}
 \eq
Taking the limit $\hbar\rightarrow 0$ and using (\ref{e057})
provides
 \beq\label{e934}
  \begin{array}{c}
  \displaystyle{
 [r_{12},m_{13}+m_{23}]=[r_{23},m_{12}+m_{13}]
 }
 \end{array}
 \eq
or (by interchanging $1\leftrightarrow2$)
 \beq\label{e935}
  \begin{array}{c}
  \displaystyle{
 [r_{12},m_{13}+m_{23}]+[r_{13},m_{12}+m_{23}]=0\,.
 }
 \end{array}
 \eq
The latter identity was used in \cite{LOZ9} for constructing the KZB
connections. More identities for $R$-matrices can be found in
\cite{LOZ11} and \cite{Z}.

\subsection{${\mathbb Z}_2$ reduction in elliptic tops}
\def\theequation{C.\arabic{equation}}
\setcounter{equation}{0}

In this paragraph we explain ${\mathbb Z}_2$ reduction (\ref{e0855})
in three ways. First, as an invariant flow of the equations of
motion
 (\ref{e04}). Second, from the geometry of the Euler-Arnold tops. And finally, as a reduction of the Lax
 equations (\ref{e040}).

 The first way is straightforward. Impose the constraints
  \beq\label{e08533}
  \begin{array}{c}
    \displaystyle{
S_\al=S_{-\al}\ \ \hbox{for}\ \hbox{all}\ \al\,.
 }
 \end{array}
 \eq
 for the
 non-relativistic top (\ref{e04})-(\ref{e0411}).
These constraints are preserved by dynamics (\ref{e0411}) because
$J_\al=J_{-\al}$ and $\kappa_{-\be,-\ga}=\kappa_{\be,\ga}$.
Therefore, the constrains are well defined.

The Euler-Arnold equations (\ref{e04}) define a flow on a coadjoint
orbit of the group $\SLN$.
 One can pass to some $\mZ_2$-invariant semi-simple subgroup $G^{inv}\subset\SLN$
and consider the Euler-Arnold equations on the coadjoint orbits in
the Lie coalgebra $(\mathfrak{g}^{inv})^*=$Lie$^*(G^{inv})$.  If the
inverse inertia tensor $J$ is also $\mZ_2$-invariant then these
orbits become invariant phase subspaces of the original phase space
(\ref{e04}).
 In what follows we use the following subgroup
\beq\label{ggr}
 G^{inv}=\left\{
 \begin{array}{lc}
   {\rm SL}(N/2+1,\mC)\times{\rm SL}(N/2-1,\mC)\times\mC^*\,, & N-{\rm\,even}\,, \\
   {\rm SL}((N+1)/2,\mC)\times{\rm SL}((N-1)/2,\mC)\times\mC^*\,, & N-{\rm\,odd}\,,
 \end{array}
 \right.
 \eq
 and $N>3$. For $N=3$, $\,G^{inv}=$SL$(2,\mC)\times\mC^*$, and for $N=2$ $\,G^{inv}=$SL$(2,\mC)$.

 For the non-relativistic tops we consider the corresponding Lie algebras. The $\mZ_2$ reduction
 is provided by the second order automorphisms $\varsigma$ of sl$(N,\mC)$.
 In terms of the generators $T_\al$ (\ref{e904}) $\varsigma$ acts as
 \beq\label{nu1}
 \varsigma\,:\,T_\al\to T_{-\al}\,.
 \eq
 Explicitly, it is defined by the conjugation by the matrix $h$
  \beq\label{e0861}
  \begin{array}{c}
    \displaystyle{
\varsigma\,:\,x\to hxh^{-1}\,,~~h=\mathcal{J}\Lambda^{-1}\,,\quad
\mathcal{J}_{ij}=\delta_{i,N-j+1}
 }
 \end{array}
 \eq
where $\Lambda$ is the one from (\ref{e903}).  It follows from
$\mathcal{J}^2=1_N$ and
  \beq\label{e0862}
  \begin{array}{c}
    \displaystyle{
\mathcal{J}\Lambda \mathcal{J} = \Lambda^{-1}\,.
 }
 \end{array}
 \eq
 that $\varsigma$ is an involution $\varsigma^2=1$.
For the matrix $Q$ from (\ref{e903}) we also have
  \beq\label{e0863}
  \begin{array}{c}
    \displaystyle{
h\,Q\,
h^{-1}=\mathcal{J}\Lambda^{-1}Q\Lambda\mathcal{J}\stackrel{(\ref{e0862})}{=}
\exp\Big(-\frac{2\pi\imath}{N}\Big)\mathcal{J}Q\mathcal{J}=Q^{-1}\,.
 }
 \end{array}
 \eq
Therefore, for the matrices $T_\al$ (\ref{e904}) we obtain
(\ref{nu1})
  \beq\label{e0864}
  \begin{array}{c}
    \displaystyle{
h\,T_{\al}\, h^{-1}=T_{-\al}\quad \hbox {for all}\ \al\,.
 }
 \end{array}
 \eq
 Therefore, the invariant subalgebra has generators $\frac12(T_\al+ T_{-\al})$, and
 by
 imposing the constraints
  \beq\label{e085}
  \begin{array}{c}
    \displaystyle{
S_\al=S_{-\al}\ \ \hbox{for}\ \hbox{all}\ \al
 }
 \end{array}
 \eq
 we come to the invariant subalgebra
 \beq\label{is}
 \mathfrak{g}^{inv}=\{\frac12\sum_\al S_\al(T_\al+ T_{-\al})\}=Lie(G^{inv})~~ (\ref{ggr})\,.
 \eq
 Since $J$ (\ref{e0400}) is also $\mZ_2$-invariant the reduction to $G^{inv}$ is consistent
 with the equations of motion.

To prove that $G^{inv}$ has the form (\ref{ggr}) we diagonalize $h$
(\ref{e0861}). The matrix $h$ has $m$ eigenvalues $\lambda=1$ and
$n$ $\lambda=-1$ $(m+n=N)$, where $m=N/2+1$ for $N$ even, and
$m=(N+1)/2$ for $N$ odd. Therefore, the subgroup of $\SLN$ commuting
with $h$
 has the form (\ref{ggr}).

As usual, to prove the integrability of the reduced system we
represent the equations of motion in the Lax form (\ref{e040}).
 Consider the Lax operator $L(z)$ (\ref{e041}). It is a meromorphic map from the complex plane $\mC$ to
 the Lie algebra sl$(N,\mC)$  satisfying fixed quasi-periodicities with respect to the shifts
  on the lattice $\mZ\oplus\tau\mZ$.
  Consider the automorphism $z\to -z$ of $\mC$. It preserves the lattice $\mZ\oplus\tau\mZ$ and in this
  way $\Sigma_\tau$.
 Consider the equivariant maps $\mC\to$sl$(N,\mC)$  with respect to the automorphisms $\varsigma$ (\ref{nu1}) and
 the automorphism $z\to -z$. It can be found that the combined actions of these automorphisms
 preserves the quasi-periodicity conditions.
 Define the Lax operator as an equivariant
 map\footnote{In fact, the Lax operator is a one-form $L(z)dz$ and the sign "$-$" in the
 r.h.s. of (\ref{e08691})
  is then absent.}
 \beq\label{e08691}
  \begin{array}{c}
    \displaystyle{
hL^{inv}(S_\al,-z)h^{-1}=-L^{inv}(S_\al,z)\,.
  }
 \end{array}
 \eq
 From (\ref{e08533}) we find the equivariant Lax operator
\beq\label{e0869}
  \begin{array}{c}
    \displaystyle{
 L^{inv}(z)=\frac12\,\sum_\al S_\al\left(\varphi_\al(z)T_\al+\varphi_{-\al}(z)T_{-\al}\right)=
 \frac12\,\sum_\al \left(S_\al+S_{-\al}\right)\varphi_\al(z)T_\al\,.
  }
 \end{array}
 \eq

The operator $M(z)$ (\ref{e041}) is map of $0$-forms  to sl$(N,\mC)$
and due to (\ref{e041}),
 (\ref{e9071}) and (\ref{e9103}) is also  the equivariant map. The equivariant maps form a Lie algebra.
Therefore, the Lax equation being reduced on the equivariant
operators $L^{inv},M^{inv}$ is equivalent to the equations of motion
on the constrained surface.

Put it differently, we can say that the set of constraints
(\ref{e08533}) is generated by involution $\varsigma$ (\ref{e0861})
acting on the Lax matrix:
  \beq\label{e086}
  \begin{array}{c}
    \displaystyle{
\varsigma (L(z,S))=h\,L(-z,S)\,h^{-1}\,.
 }
 \end{array}
 \eq
Indeed, it follows from (\ref{e0862})-(\ref{e0864}) that the action
of $\varsigma$ (\ref{e086}) on the Lax matrix (\ref{e041})  is given
as follows:
  \beq\label{e0865}
  \begin{array}{c}
    \displaystyle{
h\,L(-z,S)\,h^{-1}=\sum\limits_{\al\neq 0}T_{-\al}
S_\al\vf_\al(-z,\om_\al)=-\sum\limits_{\al\neq 0}T_{\al}
S_{-\al}\vf_\al(z,\om_\al)\,,
 }
 \end{array}
 \eq
where we used $\vf_\al(-z,\om_\al)=-\vf_{-\al}(z,-\om_\al)$. Thus,
condition
  \beq\label{e0866}
  \begin{array}{c}
    \displaystyle{
\varsigma (L(z,S))=-L(z,S)
 }
 \end{array}
 \eq
is equivalent to (\ref{e08533}).


In fact, the involution leads to decomposition
  \beq\label{e0867}
  \begin{array}{c}
    \displaystyle{
L^\pm(z,S)=\frac{1}{2}\,(L(z,S)\pm\varsigma (L(z,S))
)=\frac{1}{2}\sum\limits_{\al\neq 0}T_{\al} (S_\al\mp
S_{-\al})\vf_\al(z,\om_\al)\,.
 }
 \end{array}
 \eq
Condition (\ref{e0866}) or (\ref{e08533}) is equivalent to
$L^+(z,S)=0$, and we are left with $L^-(z,S)=L^{inv}(z)$ on the
reduced phase space.

In relativistic case we use relation to $\eta$-independent
description, i.e. from (\ref{e045}) and (\ref{e08533}) we get
  \beq\label{e087}
  \begin{array}{c}
    \displaystyle{
\frac{S_\al}{\vf_\al(\eta,\om_\al)}=\frac{S_{-\al}}{\vf_{-\al}(\eta,-\om_\al)}\,,
\ \al\neq 0
 }
 \end{array}
 \eq
and $S_0$ is not changed. Then similarly to non-relativistic case
these constraints are preserved by dynamics (\ref{e0412}).

\begin{small}

\end{small}

\end{document}